# Charge Density Wave and Ferromagnetism in Intercalated CrSBr

*Margalit L. Feuer[1†], Morgan Thinel[1,2†], Xiong Huang[2], Zhi-Hao Cui[1], Yinming Shao[2,3], Asish K. Kundu[4,5], Daniel G. Chica[1], Myung-Geun Han[4], Rohan Pokratath[6], Evan J. Telford[1,2], Jordan Cox[1], Emma York[1], Saya Okuno[1], Chun-Ying Huang[1], Owethu Bukula[1,7], Luca M. Nashabeh[2], Siyuan Qiu[2], Colin P. Nuckolls[1], Cory R. Dean[2], Simon J. L. Billinge[4,8], Xiaoyang Zhu[1], Yimei Zhu[4], Dmitri N. Basov[2], Andrew J. Millis[2,9], David R. Reichman[1], Abhay N. Pasupathy[2,4], Xavier Roy[1*], Michael E. Ziebel[1*]*

[1] Columbia University, Department of Chemistry, New York, NY, USA
[2] Columbia University, Department of Physics, New York, NY, USA
[3] Pennsylvania State University, Department of Physics, University Park, PA, USA
[4] Brookhaven National Laboratory, Condensed Matter Physics and Materials Science Department, Upton, NY, USA
[5] Brookhaven National Laboratory, National Synchrotron Light Source II, Upton, New York, USA
[6] University of Basel, Department of Chemistry, Basel, Switzerland
[7] Columbia University, Department of Mechanical Engineering, New York, NY, USA
[8] Columbia University, Department of Applied Physics and Applied Mathematics, New York, NY, USA
[9] Flatiron Institute, Center for Computational Quantum Physics, New York, NY, USA

[†] These authors contributed equally.

## Abstract

In materials with one-dimensional electronic bands, electron-electron interactions can produce intriguing quantum phenomena, including spin-charge separation and charge density waves (CDW). Most of these systems, however, are non-magnetic, motivating a search for anisotropic materials where the coupling of charge and spin may affect emergent quantum states. Here, chemical intercalation of the van der Waals magnetic semiconductor CrSBr yields $Li_{0.17(2)}(tetrahydrofuran)_{0.26(3)}CrSBr$, which possess an electronically driven quasi-1D CDW with an onset temperature above room temperature. Concurrently, electron doping increases the magnetic ordering temperature from 132 K to 200 K and switches its interlayer magnetic coupling from antiferromagnetic to ferromagnetic. The spin-polarized nature of the anisotropic bands that give rise to this CDW enforces an intrinsic coupling of charge and spin. The coexistence and interplay of ferromagnetism and charge modulation in this exfoliatable material provides a promising platform for studying tunable quantum phenomena across a range of temperatures and thicknesses.

## Introduction

Strong coupling between lattice, charge, and spin degrees of freedom underlies a diverse range of emergent phenomena in strongly correlated quantum materials, leading to complex phase diagrams where multiple ordered phases compete, cooperate, or coexist.[1] For example, mixed-valence transition metal oxides can exhibit high-temperature superconductivity, colossal magnetoresistance, charge and orbital order, and metal-insulator transitions.[2–5] The electronic and magnetic ground states of these materials are governed by a delicate interplay of localized and itinerant electrons, which are highly sensitive to both impurities and doping.[6] In three-dimensional structures, particularly non-oxide materials, synthetic control over electron density is often limited, with mobile carrier generation hindered by chemical disorder and charge localization.[6,7] Recent advances in band structure engineering of two-dimensional (2D) materials, via twisting and electrostatic gating, have enabled the realization of new correlated insulators and the study of proximate electronic phases, albeit, typically at low temperatures due to the inherently weak electron interactions.[8–12] These energy scales can be much larger in bulk 2D materials with intrinsic electron correlations, potentially enabling quantum phenomena at higher temperatures; however, while correlated phases in such materials can sometimes be influenced by doping levels accessible by electrostatic gating, these phases more typically require doping to larger electron densities.[13–15]

Intercalation of ionic or molecular species between the layers of van der Waals (vdW) materials can enable very high carrier density doping while preserving the in-plane structure.[16–18] Previous studies have identified the vdW magnetic semiconductor CrSBr, a member of the FeOCl structure family, as a correlated insulator with strong spin–phonon, phonon–exciton, and exciton–spin coupling, as well as quasi-1D electronic properties.[19–21] These couplings arise, in part, from the spin polarization and in-plane anisotropy of the two lowest energy unfilled bands of CrSBr, which we refer to as the "conduction bands" (CBs). These two bands, of primarily Cr 3*d* orbital character, originate from the two Cr atoms in the CrSBr unit cell, each forming an effectively 1D Cr−S chain along the *b* axis, with weak hybridization between adjacent chains.[22] These CBs are nearly flat along the Γ−X direction but dispersive along the Γ−Y direction (corresponding to the *a* and *b* axes, respectively) (Fig. 1a).[22,23] Shifting the Fermi level ($E_F$) into these quasi-1D, spin-polarized bands provides an intriguing avenue to probe the intrinsic coupling of lattice, charge, and spin in a strongly correlated, low-dimensional system. Here, we electron dope CrSBr through lithium-ion intercalation and investigate the 1D physics that emerge from the population of these bands. Intercalation transforms CrSBr from an A-type antiferromagnetic (AFM) semiconductor to a ferromagnetic (FM) metal, featuring a quasi-1D charge density wave (CDW) that remains stable above room temperature.

**Synthesis and structural characterization**

Previous intercalation efforts in the FeOCl structural family have primarily focused on control of local moment magnetism in insulators, such as FeOCl and CrOCl, or realizing itinerant charge carriers and superconductivity in non-magnetic semiconductors, including TiNCl and TiNBr.[24–28] Uniquely, intercalation of the magnetic semiconductor CrSBr offers an opportunity to simultaneously modify both electrons and spins. Intercalation of CrSBr is achieved through reduction of single crystals in a tetrahydrofuran (THF) solution of lithium naphthalenide, a procedure also used for intercalation of other FeOCl-type materials (Fig. 1b).[27] The reaction yields **Li-CrSBr**: partly fragmented "crystals" consisting of oriented, single crystalline domains of intercalated CrSBr (Fig. S1a-c). **Li-CrSBr** has a chemical composition of $Li_{0.17(2)}(THF)_{0.26(3)}CrSBr$, as determined by inductively coupled plasma mass spectrometry, energy dispersive X-ray spectroscopy, and $^{1}$H nuclear magnetic resonance spectroscopy (Fig. S1d-f). This level of intercalation corresponds to a doping level of ~$2.2\times10^{14}$ electrons/cm$^2$ per layer, such that $E_F$ lies within the CBs.

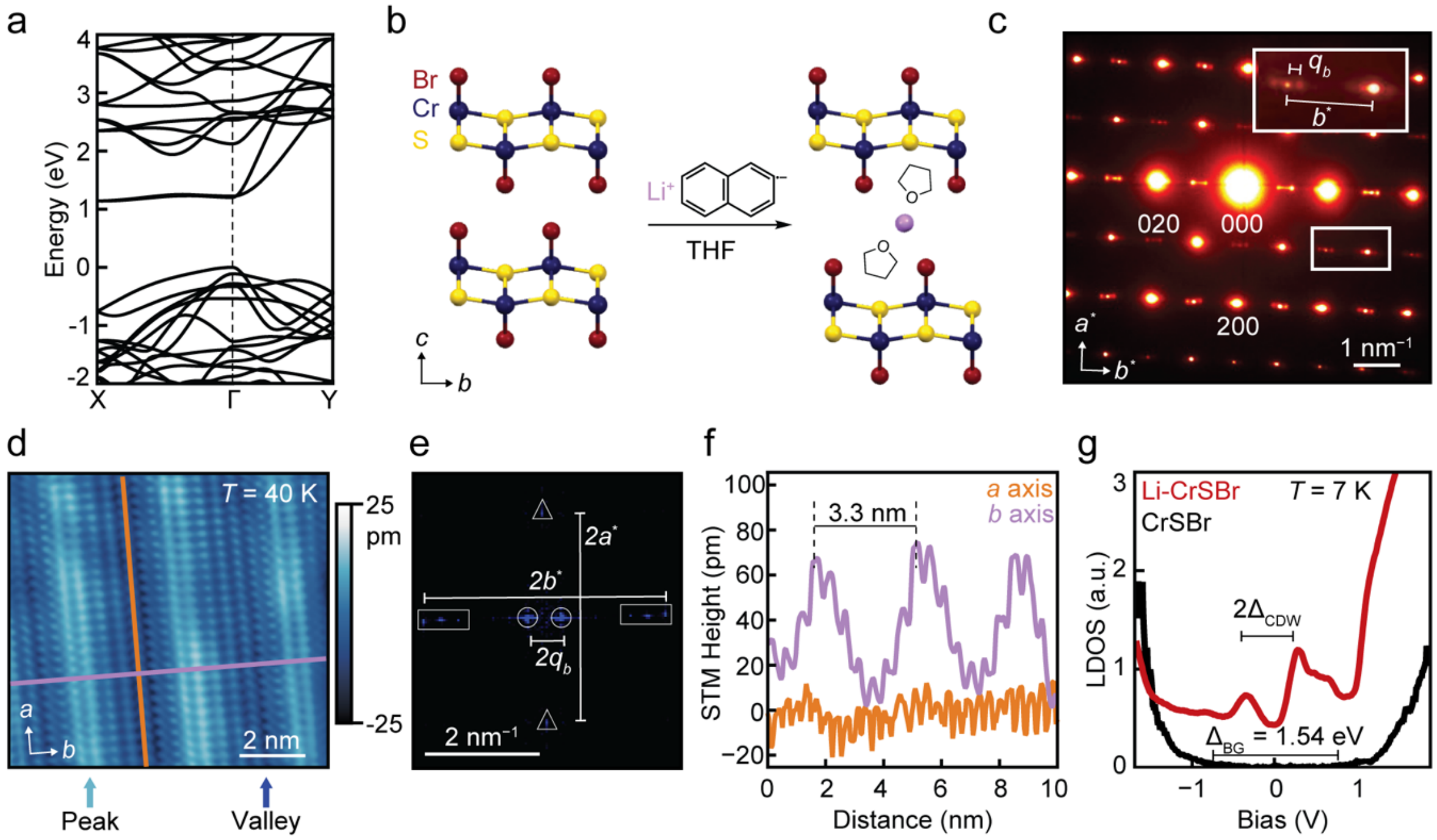

**Figure 1:** Band structure, intercalation and microscopy of CrSBr and **Li-CrSBr**.

a: Band structure of bulk CrSBr along the X−Γ−Y path calculated from DFT+U ($U$ = 4 eV, $J$ = 1 eV). b: Schematic of the intercalation of CrSBr to **Li-CrSBr**. Coordination of THF molecules to intercalated Li ions partially screens Coulombic interactions between $Li^+$ and the negatively charged CrSBr layers. Red, yellow, blue, and purple atoms correspond to Br, S, Cr, and Li, respectively. c: Room-temperature selected-area electron diffraction (SAED) of **Li-CrSBr**. The top inset box magnifies the white box beneath it to highlight the superstructure vector ($q_b$) and the $b$ axis reciprocal lattice vector ($b^*$). Slight tilting of the superlattice reflections relative to $b^*$ indicates a small $a^*$ component (see Methods). d: Scanning tunneling microscopy (STM) topography image of **Li-CrSBr** obtained in constant current mode ($V_{bias}$= 300 mV, $I_{tunneling}$ = 30 pA). The orange and purple lines show the locations for the height profiles displayed in panel (f). e: 2D fast Fourier transform (FFT) of the STM image in panel (d), with the reciprocal lattice vectors $a^*$ and $b^*$ and the superstructure ($q_b$) marked. f: Height profiles along the orange and purple lines of the STM topography shown in panel (d). Along the $b$ axis, we observe a 3.3 nm periodicity, corresponding to ~7 times the $b$ lattice constant, in addition to the atomic lattice. g: Scanning tunneling spectroscopy (STS) of pristine CrSBr (black) and **Li-CrSBr** (red). **Li-CrSBr** shows a non-zero density of states at all energies. The band gap ($\Delta_{BG}$) for CrSBr and CDW gap ($2\Delta_{CDW}$) for **Li-CrSBr** are marked.

Raman spectroscopy, X-ray pair distribution function (PDF) analysis and single crystal X-Ray diffraction (SCXRD) demonstrate that the in-plane bonding structure is preserved following intercalation (Fig. S2).[29] In-plane lattice parameters extracted from PDF and SCXRD reveal a slight change in the *a* axis lattice parameter ($|\Delta a| < 0.03$ Å) compared to pristine CrSBr, with a more significant expansion along *b* axis ($\Delta b > 0.09$ Å), consistent with the expectations for electron doping (see Discussion S1).

While the intralayer structure of CrSBr is only slightly changed by the intercalation reaction, the interlayer structure is more dramatically altered. Single crystal and powder X-ray diffraction both reveal substantial increase in the interlayer spacing, from 7.9 Å in CrSBr to more than 12 Å in **Li-CrSBr** (Fig. S2); the magnitude of this expansion is likely tied to the co-intercalation of bulky THF molecules. While some measurement-to-measurement variation is observed in the interlayer separation (see "Discussion S1" in Supporting Information), single crystal and powder X-ray diffraction measurements both suggest substantial disorder along the crystallographic *c* axis, precluding full structure refinement. Previous studies of intercalation in isostructural TiNCl revealed different shifts in the interlayer stacking depending on the size of the intercalated cation.[26] For **Li-CrSBr**, local THF disorder may similarly result in variation for both the interlayer spacing and stacking pattern. Collectively, the increase in interlayer separation and the introduction of disordered stacking suggests that the electronic structure of **Li-CrSBr** can be reasonably approximated as doped and weakly coupled monolayers of CrSBr.

Remarkably, SCXRD reveals additional reflections beyond the lattice positions (Fig. S2g). To further investigate the structure, we pursued transmission electron microscopy (TEM) and selected area electron diffraction (SAED) on thin, bulk flakes (Fig. S3). Room-temperature SAED along the [001] direction reveals sharp Bragg reflections with the expected rectangular symmetry and in-plane lattice parameters for **Li-CrSBr** (Fig. 1c). The SAED pattern shows superstructure reflections $q_b \approx b^*/7$, corresponding to a real space periodicity of ~3.4 nm. Critically, this *b* axis superstructure aligns with the high mobility direction of the CBs, suggesting a link between electron doping and the observed lattice modulation.

**Quasi-1D charge density wave**

Scanning tunneling microscopy and spectroscopy (STM/S) are widely used to investigate the surface atomic and electronic structures of materials, including CrSBr,[22,30] and to probe how superstructures affect their electronic properties. Low-temperature STM topography imaging of **Li-CrSBr** crystals reveals a unidirectional superstructure along the crystallographic *b* axis, with a periodic modulation of about seven lattice constants (~3.3 nm) that generates stripes running along the *a* axis (Fig. 1d). Fourier analysis and expanded topography images confirm the quasi-1D and long-range character of the superstructure with a superlattice wavevector, $q_b \approx b^*/7$, consistent with SAED results above (Fig. 1e, S4). Topographic line profiles along the *a* and *b* directions both show the expected modulations from the atomic lattice, with a significantly larger, wave-like modulation only along the *b* axis, highlighting the quasi-1D character of the superstructure (Fig. 1f).

Low-temperature STS measurements on **Li-CrSBr** reveal a partial gap in the local density of states (LDOS) at $E_F$, suggesting that the superstructure observed in TEM and STM can be assigned as a CDW (Fig. 1g).[31,32] By comparison, pristine CrSBr show a semiconducting bandgap ($\Delta_{BG}$ = 1.54 eV) with zero DOS at $E_F$ (Discussion S2, Fig. S5f). While the CDW periodicity is uniform across many **Li-CrSBr** samples and a gap is always observed—confirming doping-induced metallicity and a partially gapped Fermi surface in the CDW state—the magnitude of this gap ($2\Delta_{CDW}$) varies from 300 to 1000 meV in different regions (Fig. S5d,e). The gap size inhomogeneity likely arises from a mixture of local $Li^+$ and THF intercalant disorder, domain formation as well as strain (Discussion S3).[33–35] Further, consistent with room-temperature SAED measurements and the large $2\Delta_{CDW}$, room-temperature STM topography shows a superstructure with a nearly identical periodicity (~3.1 nm) to that observed at low temperatures (Fig. S5a-c).

Broadband infrared reflectivity measurements illustrate how the electronic anisotropy dominates the properties of **Li-CrSBr** (Fig. 2a). For light polarized along the *a* axis ($E \parallel a$) at $T$ = 100 K, the reflectivity of **Li-CrSBr** displays Fabry-Perot interference linked to the self-cavity effect in thin flakes.[36,37] This interference is similar to that of pristine CrSBr and indicates that, despite evidence for metallicity in STS measurements, **Li-CrSBr** remains semiconducting along the *a* axis due to the quasi-1D CBs and CDW.[38] By contrast, for $E \parallel b$, the low-energy reflectance is overall much higher than the reflectance for $E \parallel a$, in line with metallic behavior along the *b* axis.[39] We also observe a sharp increase in the reflectivity below ~1.2 eV, which we attribute to the plasma edge, indicating the presence of free carriers from $Li^+$ intercalation. At lower energies, the reflectance does not increase to unity, as expected for a simple Drude metal;[40] rather, the reflectivity along the *b* axis plateaus below 0.8 eV, dips near 0.15 eV and finally increases towards unity only very close to 0 eV. The combination of a plasma edge and a low-energy dip in the reflectivity for $E \parallel b$ is consistent with a bulk CDW and a partial gap near $E_F$, in agreement with the results above.[41]

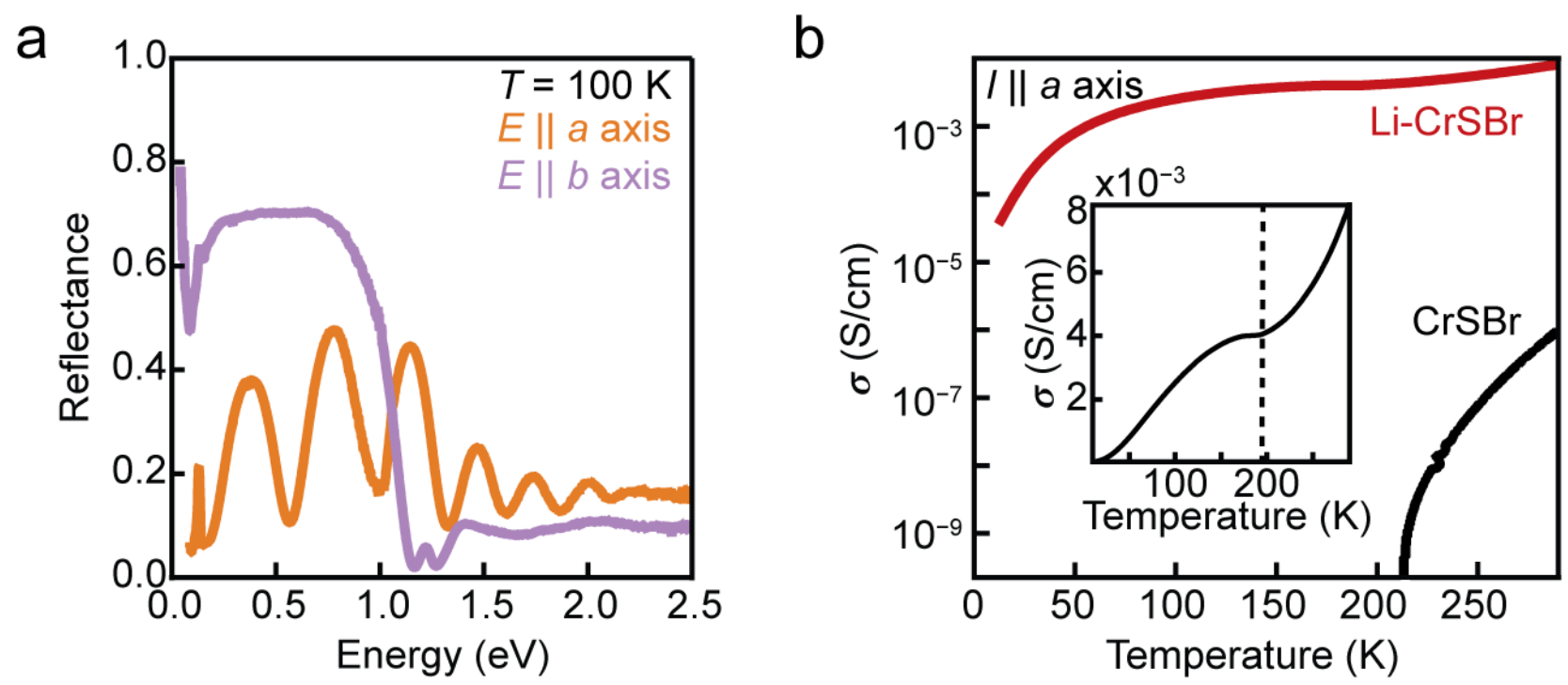

**Figure 2:** Optical anisotropy and electrical resistivity of **Li-CrSBr**.
a: Broadband reflectance spectra for **Li-CrSBr** with the light electric field (*E*) polarized along the *a* axis (orange) and the *b* axis (purple) of the crystal. The Fabry-Perot interference pattern along the *a* axis indicates semiconducting behavior. The reflectance along the *b* axis is high at low energy and exhibits a sharp drop at ~1.2 eV (plasma edge), consistent with metallic behavior along the *b* axis, b: Temperature dependence of electrical conductivity ($\sigma$) for current (*I*) flow parallel to the *a* axis of **Li-CrSBr** (red) and pristine CrSBr (black). The inset shows a broad anomaly at ~200 K in the resistivity of **Li-CrSBr** that marks the onset of magnetic order (see "Magnetic properties").

To better understand the low-energy optical feature, we performed temperature-dependent reflectivity measurements (Fig. S6a). As anticipated for a partially gapped metal, both the plasma edge and the low-energy dip broaden with increasing temperature.[41] A fit of the reflectivity data at 100 K yields a CDW gap of $2\Delta_{CDW} = 0.28$ eV (Fig. S6b). Given that the optical response is typically dominated by the smallest gap, this value aligns well with the lower bound of the gap values measured by STS.[42] We also estimate a carrier density ($n_{2D}$) ~$1.6 \times 10^{14}$ cm$^{-2}$ per layer from the extracted plasma frequency, in agreement with the doping level estimated from elemental analysis.

Electrical transport provides a secondary confirmation of semiconducting behavior along the *a* axis. Consistent with electron doping, the room temperature conductivity ($\sigma$) of **Li-CrSBr** with current flowing along the *a* axis (perpendicular to the CDW modulation) is three orders of magnitude larger than that of pristine CrSBr (Fig. 2b). Despite this doping, however, **Li-CrSBr** displays thermally activated transport, with Arrhenius-like behavior between 290 and 200 K ($E_a$ = 37 meV) and variable-range hopping behavior below 200 K (Fig. S7a-d), suggesting electron hopping between weakly coupled 1D chains is the dominant transport mechanism along the *a* axis in the CDW phase, as in CrSBr.[43]

Consistent with the STM and TEM results, the temperature dependence of the electrical conductivity of **Li-CrSBr** presents no sign of CDW suppression up to 290 K, although we note a feature at ~200 K (Fig. 2b, inset) marking the onset of magnetic order, discussed below. Furthermore, heat capacity and magnetic susceptibility measurements at higher temperatures show no CDW transition up to 390 K, signifying the CDW onset temperature ($T_{CDW}$) exceeds 390 K (Fig. S7e). Conversely, magnetic susceptibility measurements on a **Li-CrSBr** sample annealed at 473 K indicates deintercalation has occurred, establishing an effective upper bound for the thermal stability of the CDW phase (Fig. S9c). Effectively, this precludes the study of the magnetic structure across the CDW transition.

**Understanding the CDW Order**

Having identified a CDW in **Li-CrSBr**, we now turn to understanding its formation mechanism. Trivial charge order induced by an intercalant superlattice can be ruled out based on the observation of disordered surface species in STM measurements and the lack of a 3D supercell in PXRD (Discussion S4). Instead, we propose that the CDW emerges from the intrinsic properties of electron-doped CrSBr. In particular, the relationship between the CDW modulation ($q_b \approx b^*/7$) and the filling fraction of the CBs (slightly larger than $^1/_7$) points towards the importance of the electronic structure.

To evaluate the electronic contributions to the CDW in **Li-CrSBr**, we developed a minimal model using Wannier orbitals derived from the two lowest energy CBs for CrSBr to mimic the effects of electron doping (Fig. 3a, Discussion S5). Within this model, we include four main parameters: an intrachain hopping term ($t$), an intrachain Coulomb interaction ($V^{intra}$), and two interchain interactions ($V_1^{inter}, V_2^{inter}$) (Fig. 3b). Effectively, this model represents 1D, single-orbital chains, coupled via Coulomb interactions.

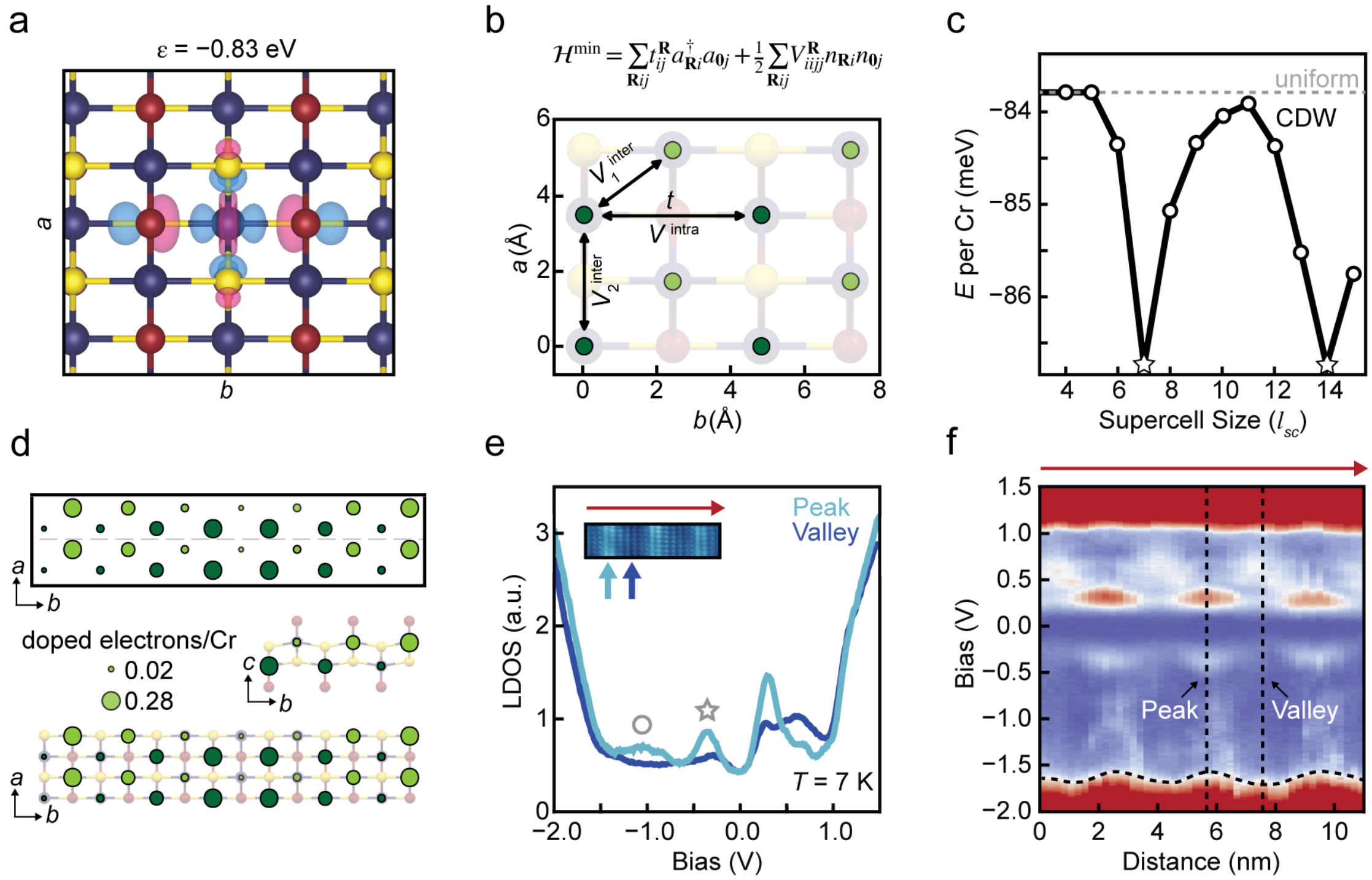

**Figure 3:** Understanding the CDW formation of **Li-CrSBr**.

a: Maximally localized Wannier functions for the $d_{z^2}$-like conduction bands of CrSBr with the on-site energy (ε). b: Schematic of the minimal model for CrSBr with the four primary model parameters: an intrachain hopping term between Cr atoms in the same sublattice (*t*), an intrachain Coulomb interaction ($V^{intra}$), and two interchain interactions ($V_1^{inter}, V_2^{inter}$). Bright (dark) green dots represent Cr atoms with higher (lower) *z*-coordinates. c: Comparison of the supercell energy per Cr atom for uniform (dashed gray) and CDW (solid black) solutions as a function of the supercell size ($l_{sc}$) for the filling fraction $\nu = {}^1/_7$. The CDW solution is energetically favorable compared to the uniform solution for $l_{sc} > 5$, and the lowest energy configurations (marked with a star) are found for commensurate CDW solutions ($l_{sc} = 7n$, where *n* is an integer). d: Schematic of the CDW solution for $l_{sc} = 7$ showing calculated electron density as circle diameter (scale shown middle-left). The middle-right and bottom panels show the projection of the CDW solution onto the **Li-CrSBr** lattice along two crystallographic orientations. e: Scanning tunneling spectra taken on a "peak" (light blue) and a "valley" (dark blue) of one CDW period. Inset shows an example STM topography image with arrows to indicate the positions of a peak and a valley. f: Scanning tunneling spectra taken over several CDW periods along the *b* axis, indicated by the red arrow above the STM inset image in (e) and mirrored at the top of (f). Vertical dashed lines indicate the peak and valley locations from STM topography; horizontal dashed line indicates the oscillating position of the valence band maximum in space. Color gradient indicates high and low local density of states.

For a filling fraction ($v$) of $^1/_7$, chosen to match the inverse of the CDW periodicity and approximate the doping level in **Li-CrSBr**, we find that a charge-modulated electronic structure is more stable than the charge-uniform electronic structure for all commensurate supercell lengths ($l_{sc}$) greater than $l_{sc} = 5$, with the most stable configuration for $l_{sc} = 7n$, where $n$ is an integer (Fig. 3c). This result strongly suggests that electron interactions drive a spontaneous breaking of the translational symmetry, giving rise to the experimentally observed CDW. The modeled charge distribution for the $l_{sc} = 7$ solution shows excess electron density oscillating between ~0.28 $e^-$/Cr and ~0.02 $e^-$/Cr along the $b$ axis (Fig. 3d). While the wave-like oscillation in the charge density is reminiscent of CDWs arising from a Peierls distortion, the amplitude of this charge modulation is large compared to those observed in conventional CDWs materials.[44–46] Effectively, this amplitude corresponds to a real-space doping modulation of the Cr atoms along the period of the CDW, which should manifest as a modulation in the valence and conduction band edges, as discussed below. Such a large CDW amplitude in this system points towards an electronic correlation-driven mechanism, as observed for charge crystals in doped spin ladder materials.[47,48]

The pattern and period of charge order in this model are in remarkable agreement with the CDW observed in STM experiments. The CDW gap calculated from mean-field theory ($2\Delta_{CDW}$ = 0.26 eV at 0 K), though an overestimate, is of similar magnitude to the gaps extracted from STS and optical measurements. We note that our model predicts a $\pi$-phase shift in the charge modulation between the neighboring Cr chains within each vdW layer, which minimizes the interchain Coulomb interactions. Due to extreme surface sensitivity, STM can only detect the Cr sites closest to the surface (bright green dots in Fig. 3d), such that the predicted phase shift is not seen in topography images. However, because of the large charge modulation predicted by this model, electronic band edges are expected to shift with local variations in doping.[44,49] This shift should be detected by a large spatial variation in the LDOS in STS measurements.

Indeed, STS shows band edges around −1.7 eV and +1.0 eV, between which there are significant variations in the LDOS measured at a "peak" or "valley" of a CDW period (Fig. 3e). Moreover, spectra measured over multiple CDW periods reveal a modulation in the energy of the valence band maximum ($E_{VBM} \approx -1.7$ eV) with the same periodicity as the CDW gap edge states ($E \approx \pm 0.3$ eV) (Fig. 3f). This modulation of the VBM provides an experimental signature for a large oscillation in the real space distribution of the "excess" electron density, as predicted by our minimal model. Notably, these large variations in the LDOS along the direction of the CDW differ significantly from what is typically observed in materials with nesting- or lattice-driven CDWs.[50–52] As such, the qualitative agreement between our purely electronic model and experiments supports a Coulomb-driven mechanism for the CDW in **Li-CrSBr**.[53]

Still, our minimal model does not fully capture our experimental observations for **Li-CrSBr**. While our model enforces two degenerate 1D bands with a perfect nesting condition at $2k_F = {}^{2\pi}/_{7}$, where $k_F$ is a Fermi wave vector, interactions between the two Cr sublattices break this degeneracy and generate finite dispersion along the Γ–X direction (Fig. S8). Experimentally, these interactions result in imperfect Fermi surface nesting, leaving the Fermi surface partially intact and explaining the observation of a non-zero DOS at $E_F$ (Fig. 1g, 3e). Moreover, our model predicts a spatial separation for electrons and holes along the CDW modulation, which would manifest as a contrast inversion upon switching STM sample bias polarity;[50] however, in our STS measurements, electron and hole DOS both reach local maxima at "peaks" of the CDW (Fig. 3e). This deviation from our purely electronic model likely indicates contributions beyond effects from Coulomb interactions. Indeed, the high temperature persistence of the CDW, unexpected for a correlation-driven mechanism, suggests that it may be partially stabilized by electron–phonon coupling, while the observation of atomic Bragg features in lattice-sensitive SAED and SCXRD measurements indicate a lattice distortion coupled to the charge modulation, distinct from what is observed in purely correlation-driven systems.[48]

**Magnetic properties**

Having established the primarily electron-driven mechanism of the quasi-1D CDW, we now turn to investigating how the addition of itinerant electrons influences the magnetic structure of **Li-CrSBr**. Pristine CrSBr is an A-type antiferromagnet below its Néel temperature ($T_N$) of 132 K: its spins align ferromagnetically along the *b* axis within each vdW layer and antiferromagnetically between adjacent layers.[30,54] The intralayer coupling is dominated by the three nearest neighbor superexchange interactions, all of which are FM.[55] Upon intercalation, these superexchange interactions should persist, with the intrachain interaction strongly enhanced by the itinerant carriers via the double-exchange mechanism.[56,57] Illustrating the enhancement, the Curie-Weiss temperature ($\theta_{CW}$) increases from 164 K for pristine CrSBr to 204 K for **Li-CrSBr** with the external magnetic field parallel to the *b* axis ($H \parallel b$) (Fig. S9a).[30] Moreover, the temperature dependence of the DC magnetic susceptibility ($\chi$) of **Li-CrSBr** displays a sharp increase below $T_C$

= 200 K, indicating the onset of FM order (Fig. 4a). As such, intercalation has two major effects on the magnetic ordering: a change in the sign of the interlayer coupling from AFM to FM, due to the combined effects of electron doping and a larger interlayer distance, and a substantial enhancement of the magnetic ordering temperature.[58,59] Importantly, chemical or thermal de-intercalation of **Li-CrSBr** restores the magnetic behavior to that of the pristine material, confirming the magnetic changes are not related to the formation of magnetic impurities upon intercalation (Fig. S9c,d).

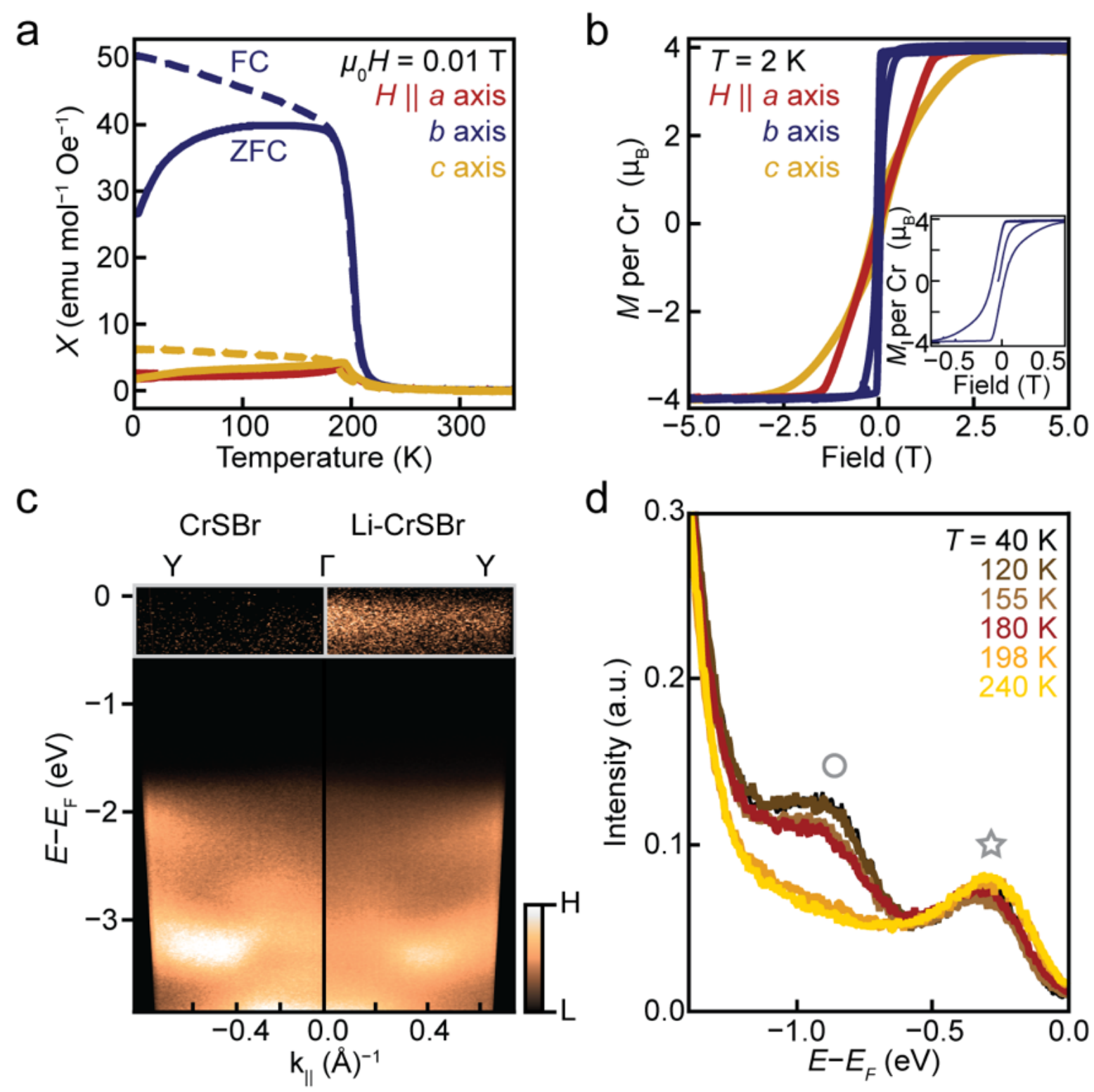

**Figure 4:** Magnetic measurements and ARPES on **Li-CrSBr**.

a: Temperature dependence of zero-field-cooled (ZFC, solid line) and field-cooled (FC, dashed line) magnetic susceptibility ($\chi$) along all three crystallographic axes. b: Field dependence of magnetization ($M$) along all three crystallographic axes. The inset shows the narrow hysteresis when $H \parallel b$ axis. c: ARPES intensity map of CrSBr (left) and **Li-CrSBr** (right) along Γ−Y at room temperature. The gray boxed areas at the top of both maps are shown with increased contrast to highlight the band with weak intensity in **Li-CrSBr** near $E_F$. The valence band edge does not significantly shift upon doping, rather the gap shrinks. d: Energy distribution curve taken at the Γ point at several temperatures on **Li-CrSBr**. The feature at $E - E_F \sim -0.3$ eV is present at all measured temperatures, while a new feature at $E - E_F \sim -0.9$ eV only emerges below $T_C$.

**Li-CrSBr** shows triaxial magnetic anisotropy, with the magnetic easy, intermediate, and hard axes aligned along the crystallographic *b, a*, and *c* axes, respectively. This anisotropy is observed in temperature-dependent susceptibility measurements, where the largest value of $\chi$ is measured for $H \parallel b$, and in field-dependent magnetization (*M*) measurements, with magnetic saturation fields $\mu_0 H_{sat}$ of 0.05 T (*b*), 1.65 T (*a*), and 2.57 T (*c*) (Fig. 4a,b). We note that the triaxial and in-plane anisotropy of CrSBr, distinctive properties among layered magnets, are retained upon intercalation.[60]

In each FM monolayer of pristine CrSBr, a pair of spin-polarized CBs share the same polarization as the VB edge.[22] Assuming rigid electron doping, the introduction of electrons into these CBs should realize a half metallic state with increased magnetization. Indeed, at 2 K, **Li-CrSBr** displays a saturation magnetization $M_{sat}$ = 4 $\mu_B$/Cr, compared to 3 $\mu_B$/Cr in CrSBr (Fig. 4b).[61] While this larger magnetization is consistent with the expected effects of electron doping, the amount of the increase is larger than expected based on the level of intercalation (expected $M_{sat} \approx 3.17$ $\mu_B$). This observation may be suggestive of a large orbital contribution to the magnetic moment, as has been observed in other magnetic CDW materials,[62] though the origin of such an orbital contribution is not obvious for **Li-CrSBr** (Discussion S6). Further measurements are needed to fully understand this behavior. Nevertheless, the increased magnetization is consistent with the occupancy of the spin-polarized bands—the same bands that are modulated by the quasi-1D CDW. Consequently, below $T_C$, where the CDW co-exists with ferromagnetism, the superlattice structure is associated with a concurrent modulation of both charge *and* spin density. Though the combination of ferromagnetic and CDW order is not unique, these phases typically manifest as a weak spin polarization of the CDW at $E_F$ due to ferromagnetic ordering of states far from $E_F$.[63] By contrast, their co-existence in **Li-CrSBr** is tied to the intrinsic spin polarization of the bands giving rise to the CDW.

Despite the large difference between $T_{CDW}$ > 390 K and $T_C$ = 200 K, this spin polarization links the charge and spin orders in **Li-CrSBr**. Given that ferromagnetism is enhanced by the double exchange interaction in **Li-CrSBr**, a likely effect of this link is that partial localization of itinerant electrons through the formation of the CDW suppresses the magnetic ordering temperature.[56] However, an intriguing question is whether the onset of ferromagnetism has effects on the CDW beyond this spin polarization. Indeed, angle-resolved photoemission spectroscopy (ARPES) measurements point to such a connection. At all measured temperatures, we observe occupied states at $E - E_F \sim -0.3$ eV in **Li-CrSBr** that are absent in CrSBr, similar to results from other doping studies (Fig. 4c).[23,64] These states align with the lower edge of the CDW gap observed in STS (Fig. 3e, denoted with a star). Upon cooling below $T_C$, a second feature emerges at $E - E_F \sim -0.9$ eV (Fig. 4d, denoted with a circle). Critically, this low temperature feature aligns with a weak but spatially modulated feature in the LDOS seen at $E \sim -0.9$ eV in STS measurements at 7 K (Fig. 3e, denoted with a circle), indicating that the new feature emerging below $T_C$ may be associated with the CDW. Further, electronic structure calculations suggest that exchange splitting of lower energy VB states cannot explain the emergence of this feature (Fig. S10). While further experiments, including higher quality ARPES and high-temperature STS, are needed to fully understand the origin of these new states, these data tentatively provide a link the CDW and magnetic order. Regardless, the spin polarization of the quasi-1D conduction band of **Li-CrSBr** creates a connection between charge and spin distinct from the typical relationship in non-magnetic 1D CDW materials;[65] this connection should enable coupled control of CDW and magnetic domains that may be applicable to spintronic devices.

For strongly correlated oxides, co-existence of magnetic and charge orders is most frequently observed in *antiferromagnetic* stripe phases.[66] Even in manganites, which possess the same $d^3/d^4$ mixed-valency as **Li-CrSBr**, FM metallic and AFM charge-ordered phases are observed in distinct compositional ranges, though co-existing ferromagnetism and charge-order have been observed in $La_{0.5}Ca_{0.5}MnO_3$ at the crossover between the two ground states.[67] By contrast, in **Li-CrSBr**, the co-existence of ferromagnetism and a CDW likely exists over a tunable range of doping. The electron-driven mechanism behind the CDW should enable longer (shorter) period CDWs at lower (higher) doping levels; indeed, in partially intercalated samples, we consistently observe a second ferromagnetic transition near $T_C^* \sim 170$ K (Fig. S9e). Given the interplay of spin and charge degrees of freedom implied from the spin polarization of the populated CBs, the lower-temperature magnetic transition may be linked to a distinct CDW phase at lower doping levels. Beyond doping, control over strain and dielectric environment in 2D devices may enable tuning of the quasi-1D charge-ordered phase.[68] Collectively, these results establish **Li-CrSBr** as a unique platform for exploring and tuning the interplay between ferromagnetism, charge order, and 1D physics.

**Outlook**

Doping correlated insulators into a metallic state is often impeded by charge localization, which has largely confined such studies to transition metal oxides. Through electron doping of the van der Waals (vdW) magnetic semiconductor CrSBr, we uncovered an unusual spin-polarized charge density wave arising from the coexistence of local-moment ferromagnetism and a quasi-1D electronic structure. The layered structure of CrSBr offer a range of tuning options, both in the bulk and at the 2D limit, allowing for exploration of its electronic and magnetic phase diagram as functions of doping, strain, and dielectric environment. Notably, the exfoliable nature of CrSBr and **Li-CrSBr** facilitates the creation of vdW heterostructures, enabling the study of magnetism and 1D physics in these materials when coupled with other electronic materials.

## Methods

### Synthesis of Li-CrSBr

CrSBr was synthesized according to previous reports using elemental sulfur, chromium bromide and chromium metal in a chemical vapor transport reaction.[55] Post-synthetic chemical intercalation reactions were carried out via a heterogeneous reaction between single-crystalline CrSBr and a lithium naphthalenide solution in tetrahydrofuran (THF). In a typical reaction, a lithium naphthalenide solution was generated by stirring a ~10mM naphthalene solution in THF over lithium metal in an argon atmosphere for 4 hours. Single crystals of CrSBr (1:1 mole ratio of Li:Cr) were added to this solution and allowed to sit overnight (without stirring), and then the solvent was decanted. This procedure was repeated two more times to reach maximum intercalation. Unless otherwise stated, all subsequent manipulations and measurements of **Li-CrSBr** were performed under an inert atmosphere with $O_2$ and $H_2O$ concentrations less than 1 ppm. Samples for spectroscopy and microscopy experiments were prepared and sealed in a $N_2$ glovebox and transferred via an inert atmosphere holder.

We note that the synthesis of **Li-CrSBr** is highly sensitive to the intercalation conditions. Attempts to intercalate CrSBr using *n*-Butyl lithium as the reductant consistently generated amorphous products (as probed by powder X-ray diffraction). Similarly, we found that a stoichiometric 1:1 ratio of Li:Cr for each intercalation step minimized the damage to crystals while allowing us to achieve homogeneous intercalation over three intercalation reactions. Smaller or larger ratios led to incomplete or inhomogeneous intercalation and degradation of single crystals, respectively.

### Exfoliation

**Li-CrSBr** crystals were mechanically exfoliated using Scotch Magic tape, and flakes were deposited onto 285 nm $SiO_2/Si^+$ substrates. This process yields flakes with estimated thicknesses from a few nanometers to several microns (Fig. S1b-c). Flake orientations are readily identify based on direction of exfoliation and the elongation of rectangular flakes along the *a* axis. For spectroscopy measurements below, thick flakes (estimated thickness >100 nm) with suitable lateral dimensions were identified by optical microscopy.

**Energy-Dispersive X-ray Spectroscopy (EDX)**

EDS measurements were performed on a ZEISS Sigma VP SEM equipped with a Bruker XFlash 6 | 30 attachment. Bulk crystals of Li-CrSBr were attached to a sample holder with conductive carbon tape. In air, samples were quickly cleaved and then placed under vacuum in the sample chamber with total air exposure limited to under 30 seconds.

**Determination of Lithium Content**

The stoichiometry of Li to Cr in **Li-CrSBr** was probed via inductively coupled plasma–optical emissions spectrometry (ICP-OES) and ICP–mass spectrometry (ICP-MS). Li and Cr standards with a concentration from 0.1 to 100 ppm were prepared by dilution of commercial standards (Sigma Aldrich TraceCert) in 3% nitric acid. Sample solutions were prepared by dissolution of single crystal and powder samples in concentrated nitric acid and subsequent dilution in 3% nitric acid to reach Li and Cr concentrations of ~1-10 ppm.

Chromium concentrations were determined using an Aglient 720 axial ICP-OES. Lithium concentration was determined using an Aglient 720 axial ICP-OES (for powder samples) or using the multiplier on a Thermo Fisher Neptune MC-ICP-MS (multi-collector ICP-MS, for low mass single crystal sample 1 below). For ICP-OES, analysis wavelengths (267.716 nm for Cr and 610.365 nm for Li) were chosen to maximize sensitivity and minimize interference. An external standard curve was used to calibrate the samples. For MC-ICP-MS, the sample was measured using a standard addition curve. For both, the standard curve range extended up to concentrations less than double that of the sample and the precision of the measurement was 1% or less.

Table 1 below shows the data for three samples. The overall Li content was found by taking the average of the Li:Cr content where the attendant error is the standard deviation.

Table 1: ICP-OES/ICP-MS data to determine Li content

| Sample # | Cr (ppm) | Li (ppm) | Li:Cr content | Notes |
|---|---|---|---|---|
| 1 | 1.38 | 0.026 | 0.14115 | 1 single crystal |
| 2 | 19.8 | 0.48 | 0.18163 | Many single crystals |
| 3 | 31 | 0.77 | 0.18610 | Many single crystals |

**X-ray photoelectron spectroscopy (XPS)**

The XPS measurements were collected using a PHI Versaprobe II spectrometer (Physical Electronics). Bulk crystals of **Li-CrSBr** were attached to the sample holder with conductive epoxy. The binding energy values of XPS peaks were calibrated using the carbon 1s peak of adventitious carbon at 284.8 eV as a reference.

**$^{1}$H Nuclear Magnetic Resonance (NMR)**

$^{1}$H NMR spectra were collected on Bruker Avance III 300 and 400SL spectrometers. As-synthesized single crystals of **Li-CrSBr** were soaked in $D_2O$ (2 mL) overnight to deintercalate Li cations and THF. The crystals were removed by filtration, and the $D_2O$ solution was mixed with a known concentration of methanol to act as an internal standard. To confirm that THF molecules remain in the material under ultrahigh vacuum conditions used for STM and ARPES measurements, single crystals of **Li-CrSBr** were evacuated ($<10^{-5}$ torr) at room temperature for 12 hours, and then subjected to the same sample preparation outlined above. Peaks corresponding to THF are visible in the $^{1}$H NMR spectra both before and after evacuation (Fig. S1f).

**X-Ray Diffraction (XRD), Pair Distribution Function (PDF), and Space Group Indexing**

Lab-based powder diffraction patterns were collected inside a $N_2$ glovebox on a Malvern Panalytical Aeris diffractometer equipped with a Cu Kα X-ray source energized to 40 kV and 15 mA and a Ni filter. The samples were ground and mounted on a Si-zero background holder, which was spun during the collection to reduce preferred orientation effects.

Single crystal X-ray diffraction data of **Li-CrSBr** was collected using a Rigaku XtaLAB Synergy-S system. In an argon-filled glovebox, freshly prepared **Li-CrSBr** material was submerged under paratone oil. The material was brought out of the glove box and a small flat ribbon of the material was mounted onto a cryoloop. The cryoloop goniometer tip was mounted onto a Kappa 4-circle goniometer and cooled to 100 K using an Oxford system cryostat. The radiation source was produced using a PhotonJet-S Cu & Mo 50W Microfocus X-ray source with the Cu anode selected. CrysalisPro software was used to control the instrument hardware for data collection, unit cell indexing, and processing the raw diffraction images into procession images.

Single crystal sample of **Li-CrSBr** were crushed in an inert atmosphere at liquid nitrogen temperatures and sealed in 1 mm polyamide Kapton tubes. Powder XRD measurements were performed at beamline 28-ID-1 at the National Synchrotron Light Source II at Brookhaven National Laboratory. Data were collected at room temperature and 130 K in rapid acquisition mode, using a Perkin Elmer digital X-ray flat panel amorphous silicon detector (2048 × 2048 pixels and 200 × 200 μm pixel size). The sample to detector distance was 239 mm. The incident wavelength of the X-rays was $\lambda = 0.1665$ Å. Calibration of the experimental setup was performed using a Ni standard.

Raw 2D data were corrected for geometrical effects and polarization, then azimuthally integrated to produce 1D scattering intensities versus the magnitude of the momentum transfer Q (where $Q = 4\pi\sin\theta/\lambda$ for elastic scattering) using pyFAI and xpdtools.[69,70] For PDF analysis, the program xPDFsuite with PDFgetX3 was used to perform the background subtraction, further corrections, normalization to obtain the reduced total scattering structure function F(Q), and Fourier transformation to obtain the pair distribution function, G(r).[71,72] For data reduction, the following parameters were used after proper background subtraction: $Q_{min} = 0.8$ Å$^{-1}$, $Q_{max} = 22$ Å$^{-1}$, $R_{poly} = 0.9$ Å. Modeling and fitting were carried out using Diffpy-CMI.[73] For unit cell analysis of the synchrotron PXRD data, GSAS-II software was used to perform space group indexing and Pawley refinement of the unit cell parameters (Fig. S2d).[74]

**Physical Property Measurements**

Magnetometry, heat capacity, and electrical transport measurements were performed on a Quantum Design Physical Property Measurement System (PPMS) DynaCool or Quantum Design VersaLab. For magnetometry, single crystals were weighed and encapsulated between two quartz plates with super glue to keep the sample air-free during sample loading. For measurements with the magnetic field applied along the crystallographic *a* or *b* axes, the sample was attached to a quartz paddle with GE varnish. For measurements with the magnetic field applied along the crystallographic *c* axis, the sample was glued on top of a quartz plug fitted within a brass holder. The diamagnetic contribution of the sample holder was determined from the negative slope of low-temperature measurements of moment as a function of magnetic field, and a diamagnetic correction was applied to the raw measured moment of all data.

For heat capacity measurements, a known mass of **Li-CrSBr** powder was pelletized with a small mass percentage of Cu powder to improve thermal equilibration. The samples were mounted with Apiezon N grease for measurements in the range 2–300 K and Apiezon H grease for measurements in the range 280–389 K (instrument limitations prevented measurement between 389 and 400 K). For all measurements, a piece of gold foil (0.0004 inches thick) was used to encapsulate and protect samples during the transfer into the PPMS. The contributions of grease and the gold foil were measured and subtracted from the raw sample heat capacity; the contribution of copper metal was subtracted from the raw sample heat capacity as well.

For electrical transport measurements, electrical connections to the sample were made with silver paint (Dupont 4929N) and 25 μm diameter 99.99% gold wire. For pristine CrSBr, crystals were cleaved to obtain a mirror surface prior to device preparation. The longitudinal resistance of CrSBr was measured in a two-terminal configuration using an SRS830 lock-in amplifier to source voltage and measure current using a 17.777 Hz reference frequency. For **Li-CrSBr**, crystals were encapsulated in Apiezon N vacuum grease to protect the sample from air during loading. The resistance of three-contact devices was measured with an AC current of 10 μA applied along the crystallographic *a* axis. The fragility and air/temperature sensitivity of **Li-CrSBr** crystals have so far hindered transport measurements along the *b* axis (i.e. along the CDW modulation). Bulk, two contact devices measured with current along the *b* axis show anomalously large resistances (>25kΩ), suggesting poor contacts or resistance dominated by grain boundaries. Future measurements on exfoliated flakes may enable more detailed study of the *b*-axis transport properties.

**Scanning Tunneling Microscopy and Spectroscopy (STM/STS)**

Low-temperature STM data were obtained using a home-made variable-temperature STM. Room temperature data were obtained using an Omicron VT-STM. Bulk crystals of **Li-CrSBr** were affixed to metallic holders with silver epoxy, cleaved with Scotch tape, and then cleaved *in situ* under ultra-high vacuum using Kapton tape prior to STM measurements. Chemically etched tungsten tips were annealed and calibrated on a Au (111) or Cu (111) surface. Topographic images were acquired in constant-current mode. Spectroscopic data were taken by the standard lock-in technique with bias voltage applied to the sample with a fixed tip-sample distance. Topographic image visualization was analyzed using a free and open-source software, Gwyddion.[75]

For low-temperature data on thin flakes of pristine CrSBr (<200 nm), a conducting Au-film-coated mica served as the substrate to form a closed tunneling circuit. To prepare samples, a commercial Au (111) surface deposited on mica (Phasis) was pressed to the surface of bulk CrSBr crystals. This process yields thin flakes of pristine CrSBr, which were subsequently measured as described above.

**Transmission Electron Microscopy (TEM)**

**Li-CrSBr** flakes were exfoliated and directly transferred to a TEM grid with 1 μm diameter holes in commercial silicon nitride membranes (Ted Pella). Electron diffraction patterns were collected on a Jeol 2100F electron microscope with a 200 keV electron beam. The absence of an amorphous ring in collected electron diffraction patterns confirms that the sample does not degrade during loading or measurement.

We note a slight tilt in the superlattice reflections relative to the $a^*$ with $q_a \approx a^*/50$ and a real space periodicity of ~17.4 nm. This small *a* axis contribution is not found in other spectroscopic techniques (including STM and optical reflectivity) and may be related to buckling of the CrSBr layers in the intercalated material.

**Raman Spectroscopy**

Raman spectroscopy on exfoliated flakes of **Li-CrSBr** was performed under a $N_2$ atmosphere in a Renishaw inVia Raman microscope using a 532 nm wavelength laser. A 20× objective was used with a laser spot size of 5–7.5 μm. A laser power of ~25 mW was used with a grating of 1800 l mm$^{-1}$ for all spectra. An acquisition time of 120 s was used for each measurement. For each flake, spectra were acquired with the laser polarized along the *a* or *b* axes. Five independent spectra were averaged for each polarization direction after subtracting a dark background.

**ARPES**

Bulk crystals of **Li-CrSBr** were affixed to Cu foil with silver epoxy and cleaved with Scotch tape. Samples were then clamped to the sample holder and cleaved with the Kapton tape in the ARPES preparation chamber with a base pressure of 1 × 10$^{-9}$ torr. The ARPES experiments were carried out using a Scienta SES-R4000 electron spectrometer with monochromatized HeI (21.22 eV) radiation (VUV-5k) with a base pressure of 2 × 10$^{-10}$ torr. The total experimental energy resolution were ～20 meV and ～100 meV for low- and room-temperature measurements, respectively.

### Broadband reflectance measurements

The broadband reflectance spectra of exfoliated **Li-CrSBr** flakes were measured using a Hyperion 2000 microscope coupled with a Bruker Fourier-transform infrared spectrometer (Vertex 80V). A Globar light source was used covering the energy range of 0.04 to 0.8 eV, and a tungsten halogen lamp was used as a light source covering an energy range of 0.5 to ~2.5 eV. Linear polarizers were used to control the polarizations of light, which was focused on the sample using a 15× objective and the aperture size was set to be smaller than the sample dimensions (37 μm × 31 μm). The reflectance spectra are normalized to an optically thick silver mirror. A 4.2 K Si Bolometer, a mercury–cadmium–telluride (MCT) detector, and a silicon detector were used for the far-infrared, mid-infrared, and visible range, respectively. Temperature dependent measurements were carried out using the same optical set-up with the sample mounted inside a Helium flow cryostat with a base temperature of ~5 K (Oxford MicrostatHe). For data in Fig. 2, an exfoliated flake with a thickness of 542 nm was used, as determined by Fabry-Perot interference.

### Computational Modeling

The hopping parameters were extracted from first-principles density functional theory calculations, and the screened Coulomb interactions were evaluated using the constrained random phase approximation. We employed the spin-polarized Perdew-Burke-Ernzerhof (PBE) functional for the ferromagnetic electronic structure of monolayer CrSBr (based on the experimental bulk structure with $a$ = 3.512 Å, $b$ = 4.746 Å, and 14 Å additional vacuum in the z-direction).[61,76] DFT calculations were performed using the VASP package[77,78] with a plane-wave basis cutoff of 500 eV. The Brillouin zone was sampled using a Γ-centered 18×14×1 $k$-point mesh. The lowest four conduction bands were localized using the Wannier90 code,[79] with a disentanglement window of −2.15 to 1.0 eV and Cr-$e_g$ projections. The cRPA was performed using a 2×2×1 cell, excluding 16 target Wannier states from the dielectric screening.[80,81]

The Hartree-Fock simulations on the model were carried out using the libDMET[82–84] and PySCF packages.[85,86] We used a $1\times1_{sc}\times1$ supercell with a Γ-centered 21×20×1 $k$-point mesh. We considered uniform, CDW (with different periodicities), and random guesses to ensure finding the global ground state.

**Corresponding Authors**

*xr2114@columbia.edu

*mez2127@columbia.edu

**Acknowledgments**

This work was primarily supported by the NSF through the Columbia University Materials Research Science and Engineering Center (MRSEC) on Precision-Assembled Quantum Materials DMR-2011738 and CAREER award DMR-1751949. Magnetic measurements were supported as part of Programmable Quantum Materials, an Energy Frontier Research Center funded by the DOE, Office of Science, Basic Energy Sciences (Award DE-SC0019443). Scanning tunneling microscopy measurements were supported in part by the US Air Force Office of Scientific Research (AFOSR) under award number FA9550-22-1-0389. This research used resources of the Condensed Matter and Materials Science Division and of beamline 28-ID-1 of the National Synchrotron Light Source II, a U.S. Department of Energy (DOE) Office of Science User Facility operated for the DOE Office of Science by Brookhaven National Laboratory under Contract No. DE-SC0012704. PDF analysis in the Billinge group was supported by U.S. Department of Energy, Office of Science, Office of Basic Energy Sciences (DOE-BES) under contract No. DE-SC0024141. We thank Maya Nair for her help with the XPS run at Advanced Science Research Center of City University of New York. We thank Louise Bolge for her help with ICP-MS. M.L.F. is supported by the NSF Graduate Research Fellowship Program (DGE-2036197). The PPMS used to perform magnetic susceptibility measurements was purchased with financial support from the NSF through a supplement to award DMR-1751949. The Columbia University Shared Materials Characterization Laboratory (SMCL) was used extensively for this research. The authors are grateful to Columbia University for the support of this facility. The Flatiron Institute is a division of the Simons foundation.

## Author contributions

M.L.F synthetized **Li-CrSBr** samples, performed magnetic measurements and electronic transport with E.J.T. and O.B. supporting. M.T. and X.H. performed scanning tunneling microscopy and spectroscopy with L.M.N. supporting. Z.-H. C. performed the computational modeling under the supervision of D.R.R and A.J.M. Y.S. and S.Q. performed optical reflectivity measurements under the supervision of D.N.B. A.K.K. performed ARPES measurements. D.G.C synthesized pristine CrSBr crystals. M.-G.H. and Y.Z. performed the transmission electron microscopy imaging. R.P. performed pair distribution function analysis and synchrotron powder X-ray diffraction under the supervision of S. J. L. B. Structural analysis was performed by J.C., E.Y., S.O and C.-Y. H. under the supervision of C.P.N. and X.-Y.Z. A.J.M., D.R.R, C.R.D., A.N.P., X.R. and M.E.Z. supervised the project. M.L.F., M.T., X.R. and M.E.Z. wrote the manuscript with input from all authors.

## Competing interests

The authors declare no competing interests.

**Supplementary Information:**

# Charge Density Wave and Ferromagnetism in Intercalated CrSBr

*Margalit L. Feuer*[1†], *Morgan Thinel*[1,2†], *Xiong Huang*[2], *Zhi-Hao Cui*[1], *Yinming Shao*[2,3], *Asish K. Kundu*[4,5], *Daniel G. Chica*[1], *Myung-Geun Han*[4], *Rohan Pokratath*[6], *Evan J. Telford*[1,2], *Jordan Cox*[1], *Emma York*[1], *Saya Okuno*[1], *Chun-Ying Huang*[1], *Owethu Bukula*[1,7], *Luca M. Nashabeh*[2], *Siyuan Qiu*[2], *Colin P. Nuckolls*[1], *Cory R. Dean*[2], *Simon J. L. Billinge*[4,8], *Xiaoyang Zhu*[1], *Yimei Zhu*[4], *Dmitri N. Basov*[2], *Andrew J. Millis*[2,9], *David R. Reichman*[1], *Abhay N. Pasupathy*[2,4], *Xavier Roy*[1*], *Michael E. Ziebel*[1*]

[1] Columbia University, Department of Chemistry, New York, NY, USA
[2] Columbia University, Department of Physics, New York, NY, USA
[3] Pennsylvania State University, Department of Physics, University Park, PA, USA
[4] Brookhaven National Laboratory, Condensed Matter Physics and Materials Science Department, Upton, NY, USA
[5] Brookhaven National Laboratory, National Synchrotron Light Source II, Upton, New York, USA
[6] University of Basel, Department of Chemistry, Basel, Switzerland
[7] Columbia University, Department of Mechanical Engineering, New York, NY, USA
[8] Columbia University, Department of Applied Physics and Applied Mathematics, New York, NY, USA
[9] Flatiron Institute, Center for Computational Quantum Physics, New York, NY, USA

[†] These authors contributed equally.

# Table of Contents

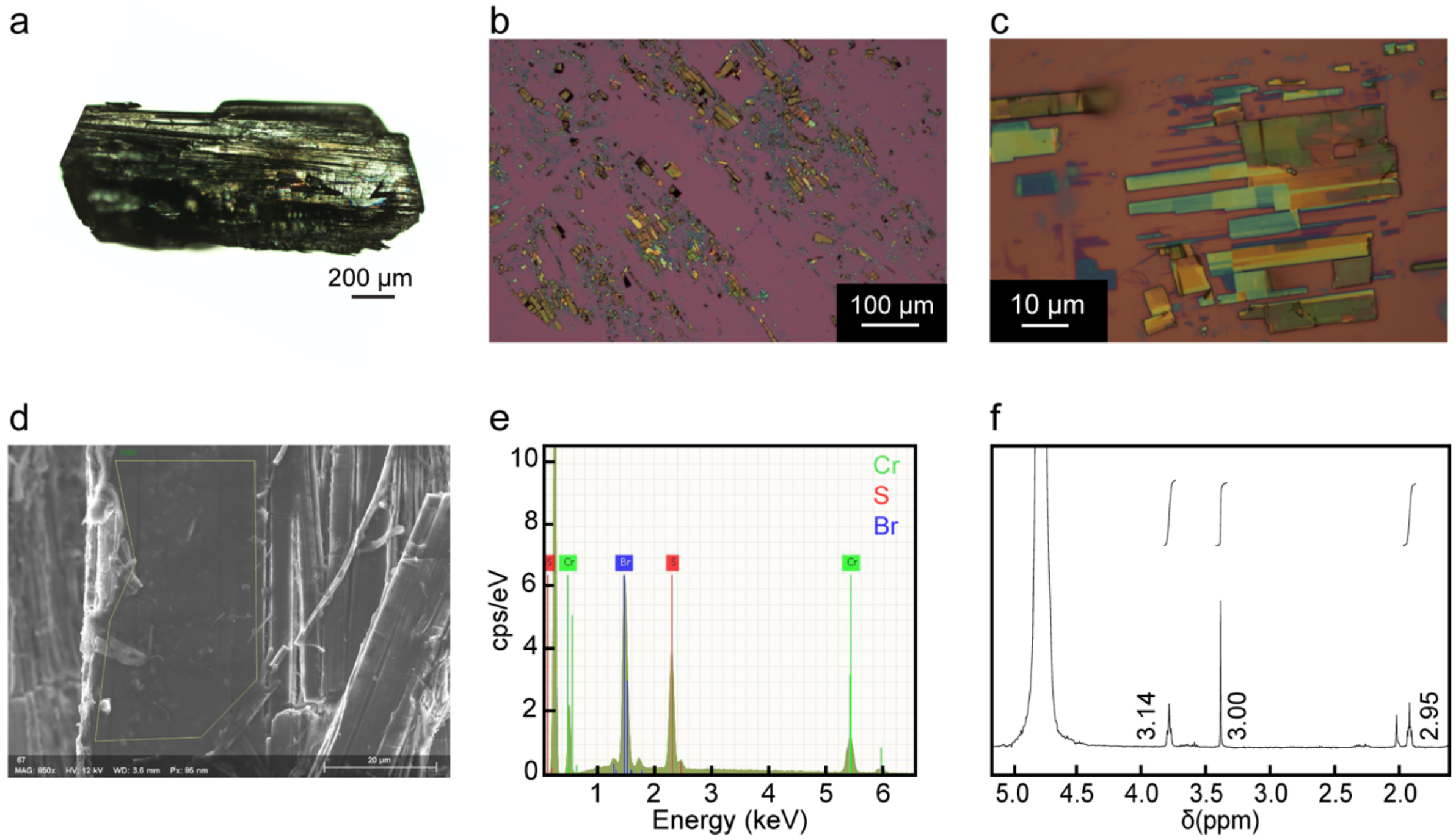

**Figure S1: Microscope images, SEM-EDS, and NMR analysis of Li-CrSBr.**
a: Optical microscope image of a **Li-CrSBr** crystal to show the size and surface roughness. b,c: Flakes of **Li-CrSBr** after mechanical exfoliation with Scotch magic tape on $SiO_2/Si^+$ chip. Colors corresponds to a range of thicknesses of the flakes. Typical lateral dimensions are on the order of 10–50 µm with the long axis corresponding to the crystallographic *a* axis. d,e: SEM image and emitted X-ray spectrum from EDS show the retention of the 1:1:1 stoichiometry of Cr, S, and Br. This indicates that the reaction with $Li(C_{10}H_8)$ proceeds almost exclusively through intercalation of lithium, rather than through abstraction of bromide or sulfide. f: $^{1}H$ NMR spectrum of the supernatant when **Li-CrSBr** was submerged in $D_2O$ with known concentration of methanol used as an internal standard, which shows peaks corresponding to THF. Peaks: δ 1.19 (t, 2H), 3.36 (s, 3H), and 3.78 (t, 2H). Integration is labeled next to the peak. These co-intercalated THF molecules likely coordinate to and electrically screen the $Li^+$ charges, as has been previously observed in the isostructural TiNBr.[1] THF is still observed in **Li-CrSBr** following prolonged evacuation at room temperature, as seen by NMR as well. The large change in the interlayer spacing required to accommodate THF likely contributes to the fragmentation and surface roughening of CrSBr crystals during the intercalation reaction.

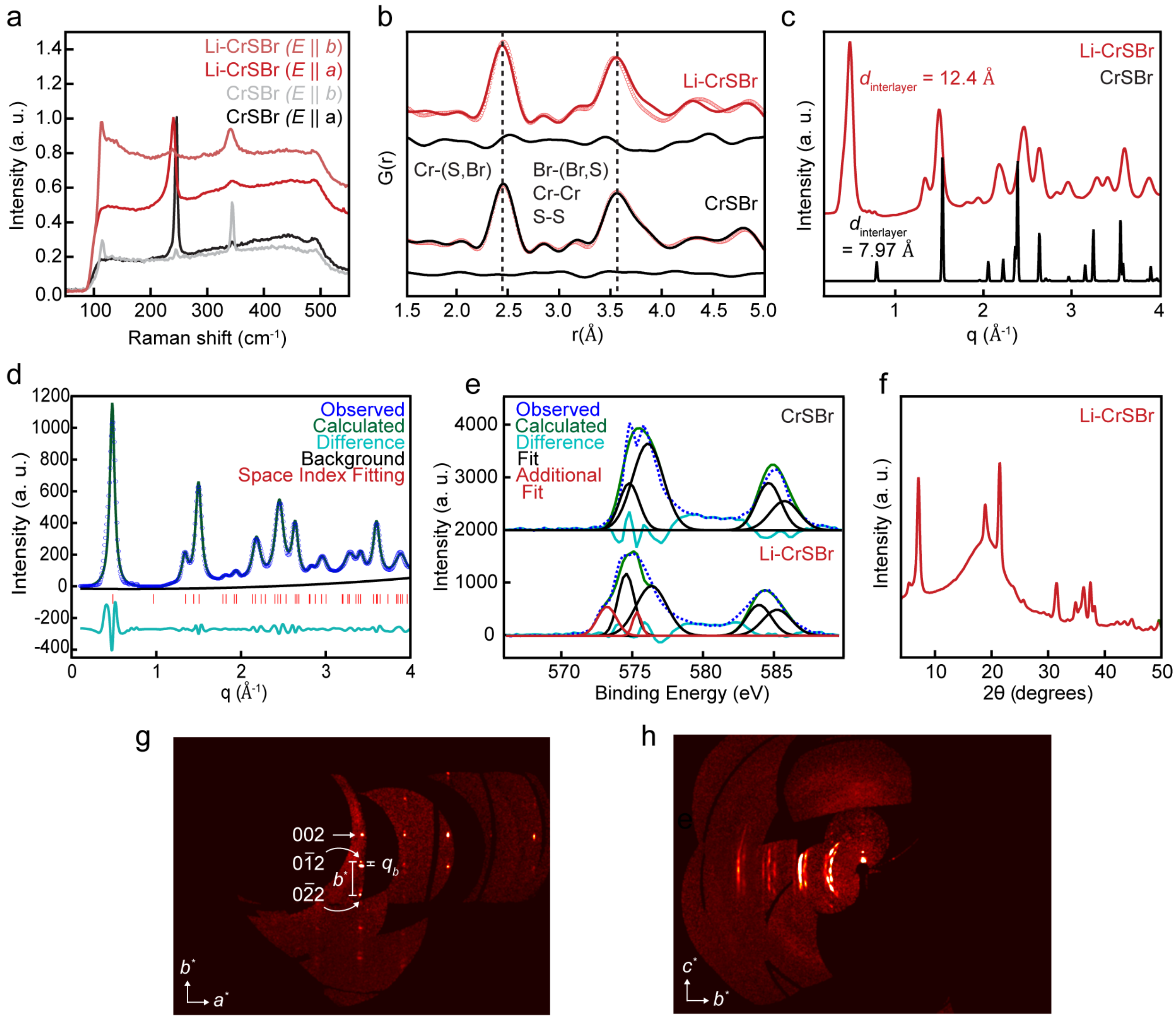

**Figure S2: Structural analysis of Li-CrSBr.**

a: Raman spectroscopy of **Li-CrSBr** and CrSBr with the incident laser polarized along both the *a* and *b* axes. All three Raman modes ($A_g^1$, $A_g^2$, and $A_g^3$ from left to right) are retained upon intercalation with analogous dependence on incident light polarization and a small amount of phonon softening. This softening is likely related to the change in interlayer spacing.[2] b: Pair distribution function (PDF) analysis of short-range bonds of **Li-CrSBr** and CrSBr with data (open circles) and fit (solid line). Significant peaks are labeled with the bonds they represent and a dashed black line. The shoulder in the PDF of CrSBr at 3.8 Å, which corresponds to the interlayer Br–Br separation, is absent in that of **Li-CrSBr**, indicating a change of the interlayer spacing upon co-intercalation of $Li^+$ and bulky THF. Refined parameters are shown in Table S1 below. c: Synchrotron PXRD of **Li-CrSBr** and CrSBr (ICSD - coll. code 69659)[3] with $d_{interlayer}$ marked to show the increase in interlayer spacing. d: Unit cell refinement of the synchrotron PXRD of **Li-CrSBr** with lattice parameters in Table S2 below. e: XPS spectra of **Li-CrSBr** and CrSBr of Cr $2p_{3/2}$ and $2p_{3/2}$ peaks. The red peaks in **Li-CrSBr** shows the reduction of Cr after intercalation. f: X-ray diffraction of a single-crystal of **Li-CrSBr** in Gandolfi mode at 100 K. There are two peaks at 5.41° ($d$ = 16.3 Å) and 7.10° ($d$ = 12.4 Å), closely matching two phases that are seen in SCXRD and PXRD, respectively. g,h: Single crystal X-ray diffraction procession images of (*h*, *k*, 2) (d) and (0, *k*, *l*) (e) with a space group of $P\overline{1}$. See the full unit cell parameters in Table S3.

**Table S1: Refined in-plane parameters from PDF analysis after fitting pristine CrSBr and Li-CrSBr at room temperature.**

| | Scale | *a* (Å) | *b* (Å) | $U_{iso}$ (Cr) | $U_{iso}$ (S) | $U_{iso}$ (Br) | $R_w$ (%) |
|---|---|---|---|---|---|---|---|
| CrSBr | 0.37 | 3.50 | 4.78 | 0.002 | 0.012 | 0.014 | 9 |
| **Li-CrSBr** | 0.047 | 3.49 | 4.87 | 0.002 | 0.012 | 0.034 | 22 |

**Table S2: Powder X-ray diffraction unit cell refinement at room temperature.**

| *a* (Å) | *b* (Å) | *c* (Å) | M20 | X20 | Space Group Indexed |
|---|---|---|---|---|---|
| 3.50 | 4.78 | 26.26 | 22.50 | 0 | *Immm* |

**Table S3: Single crystal X-ray diffraction unit cell parameters of Li-CrSBr at 100 K.**

| *a* (Å) | *b* (Å) | *c* (Å) | $\alpha$ (°) | $\beta$ (°) | $\gamma$ (°) | Volume ($Å^3$) |
|---|---|---|---|---|---|---|
| 3.533 | 4.878 | 16.70 | 91.2 | 84.9 | 90.7 | 287 |

## Discussion S1: Lattice parameter changes

### In-plane lattice parameters:

In previous studies on intercalation of TiNCl, which is isostructural to CrSBr, lithium-ion intercalation resulted in an isotropic expansion of the in-plane lattice parameters, consistent with an increase in the ionic radius of Ti upon reduction.[4] For CrSBr, while the overall area of the in-plane unit cell expands, the unit cell expansion is anisotropic. A significant expansion (~2%) is observed along *b* across both PDF and SCXRD. While slight differences in the *a* axis lattice parameter are found for room temperature X-ray PDF analysis compared to 100 K SCXRD (see tables above), the lattice change along *a* is markedly smaller than along *b*. The observed anisotropic expansion along *b* is consistent with the quasi-1D electronic structure of CrSBr: The conduction bands are derived from a σ* orbital oriented along the *b* axis;[5] population of these orbitals upon electron doping should primarily result in an expansion along this direction. As such, in-plane structural changes upon intercalation are as expected for rigid electron doping of CrSBr.

### Out-of-plane lattice parameters:

The interlayer spacing is, by all measurements, (1) larger than in pristine CrSBr and (2) significantly disordered. Regarding the magnitude interlayer spacing, PXRD of ground material at room temperature presents an interlayer spacing of 12.4 Å, whereas SCXRD of unground material at 100 K reveals an interlayer spacing of 16.70 Å. We note, however, that averaged diffraction of unground single crystals suggests a primary phase with interlayer spacing of ~12.4 Å and a minority phase yielding single crystalline spots with an interlayer spacing of ~16.3 Å (Fig. S2f). This could reflect changes in the amount of co-intercalated THF or the orientation of THF/Li-THF complexes between the layers. Regardless, the interlayer distance is significantly expanded from the pristine material ($c$ = 7.965 Å), which suggests that the intercalated material is a functionally 2D crystal from an electronic and magnetic perspective.

### Space group determination:

Unit cell refinement of PXRD (Fig. S2d) identifies the space group *Immm* for **Li-CrSBr** (compared to *Pmmn* for CrSBr). This symmetry change signifies that the layers of **Li-CrSBr** are shifted by a half unit cell along both *a* and *b* axes when compared to CrSBr, doubling the unit cell along the *c* axis. Similar structural transformations occur upon intercalation of isostructural titanium nitride halides, where the interlayer shift creates a rectangular prismatic cage of halide ions to host intercalants.[1,4] In contrast, the minority phase detected in SCXRD experiments yields a triclinic unit cell (space group $P\overline{1}$), with a small shift between adjacent CrSBr layers. However, smearing of *c* axis peaks in SCXRD measurements are indicative of substantial stacking disorder. Collectively, these data support a mixture of stacking patterns within **Li-CrSBr**, which is unsurprising for a low-temperature intercalation reaction with a large increase in the interlayer spacing. Attempts to better characterize the interlayer stacking with cross-sectional TEM measurements were hindered by the air-sensitivity of **Li-CrSBr**. As such, we believe the *c*-axis structure of **Li-CrSBr** is best represented as random stacking, though we cannot rule out domains with short-range stacking order.

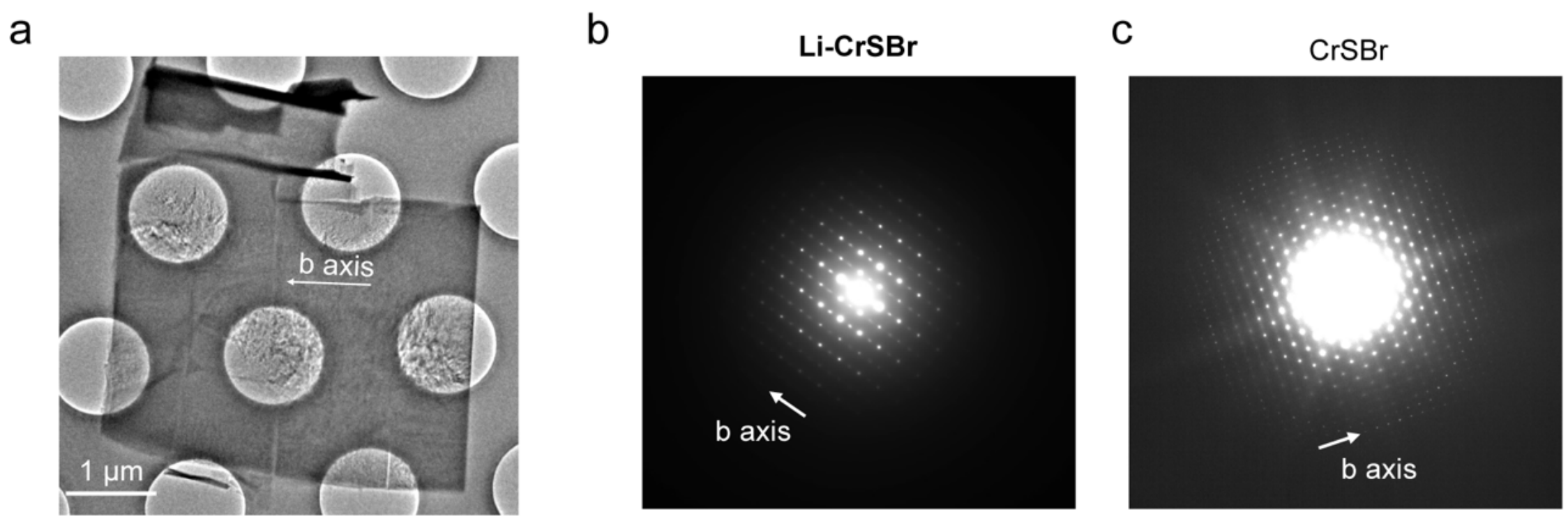

**Figure S3: Transmission electron microscopy and electron diffraction patterns.**
a: TEM image of **Li-CrSBr** on a silicon nitride grid. b,c: Electron diffraction pattern of **Li-CrSBr** (b) and CrSBr (c) to show a first-order Laue zone reflection in CrSBr and the absence of the first-order Laue zone reflection in **Li-CrSBr**. The presence of a first-order Laue zone reflection implies 3D order along the stacking axis. In contrast, panel (b) indicates the lack of crystal ordering along the *c* axis and explains the absence of superlattice reflections in PXRD.

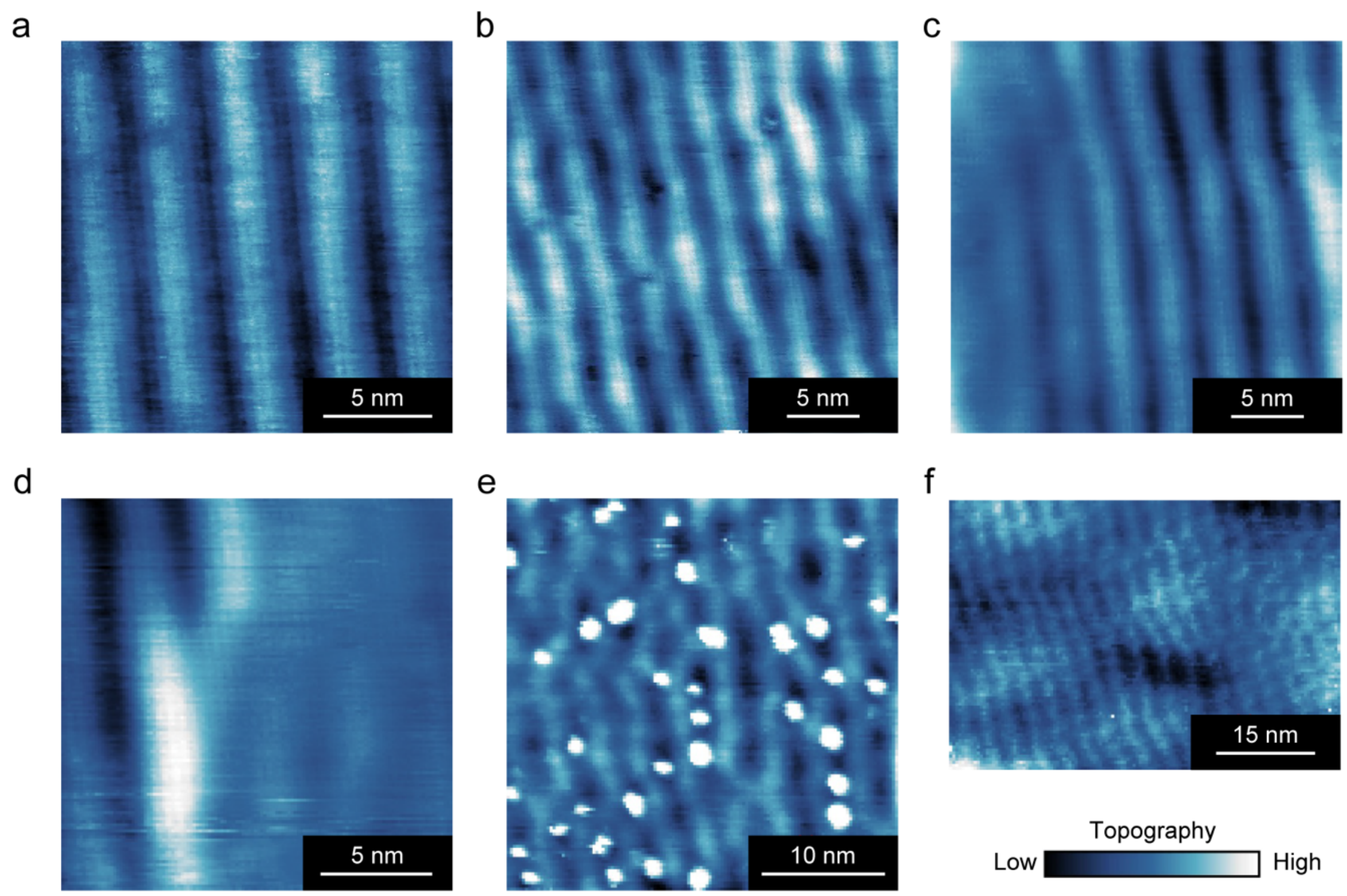

**Figure S4: Representative STM topography images on Li-CrSBr samples.**
The STM topography above are taken in different areas of **Li-CrSBr** crystals at 7K. Despite the variations in STM topography shown here, the variations in gap sizes shown in Fig. S5 do not correlate with changes in CDW periodicity. a: STM topography showing a clean surface with a quasi-commensurate CDW period. b,c: Disorder (likely due to defects and dislocations) lead to incommensurate CDW periods that manifest as stripes that are no longer parallel along the *a* axis. d: A domain boundary suppresses the magnitude of the CDW induced topographic distortions. e: An area that is covered in adatoms or molecules. Under the assumption that these are THF molecules co-intercalated with $Li^+$, this provides evidence for non-uniform distribution of the $Li^+$ ions despite a uniform long-range CDW observed in clean areas. f: Highly disordered regions can show complex patterns that lack the stripe coherence observed in clean areas.

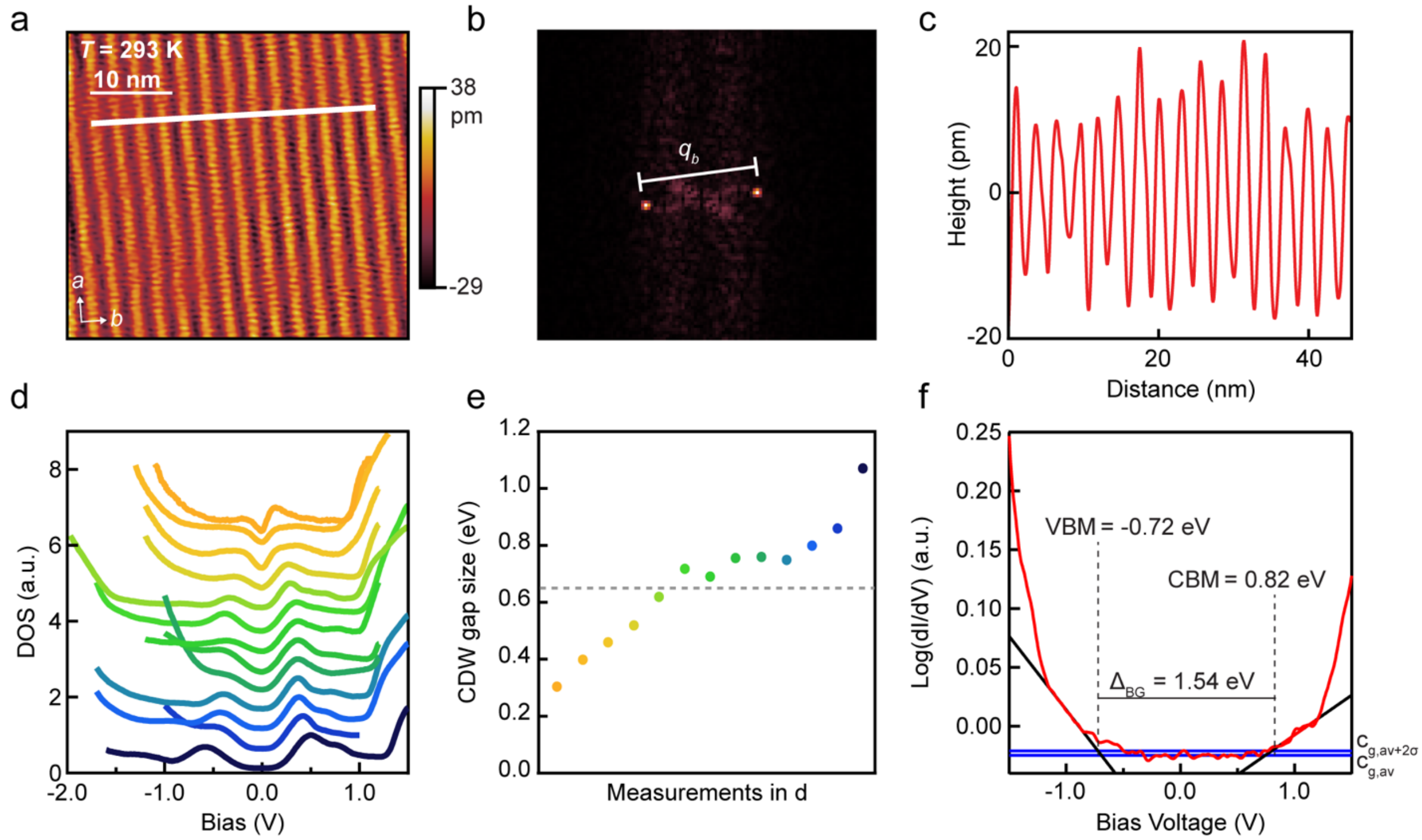

**Figure S5: Room-temperature STM with analysis, extracted CDW band gaps of Li-CrSBr, and CrSBr band gap.**

a: Room-temperature STM topography image obtained in constant current mode ($V_{bias}$= −100 mV, $I_{tunneling}$ = 50 pA). We note that STS spectra could not be resolved at room temperature. b: The FFT image of (a) with the $q$ vector of the CDW marked. c: The topographic line profile along the $b$ axis marked by the white line in (a). d, e: STS spectra at different sample surfaces (d) with the CDW gap value plotted (e). The gray dashed line indicates the value of 650 meV given in the main text. Data was taken at 7 K. Discussion of the gap size determination is below. f: Determination of the semiconducting gap in pristine CrSBr with further discussion below.

**Discussion S2: Pristine CrSBr gap size**

In analyzing the STS data on **Li-CrSBr**, it is useful to compare to the behavior of the pristine material. The magnitude of the band gap of CrSBr has recently received substantial theoretical and experimental attention. Here, we evaluate the band gap at low temperature using relatively thick samples of CrSBr exfoliated onto a Au (111) surface.

Due to the wide variation in signal strength near a band edge, using the logarithm of the dI/dV curves describe a rigorous method for quantitatively determining the band gap of a 2D semiconductor.[6] As shown in Fig. S5f, we first calculate the mean ($C_{g,av.}$) and standard deviation ($\sigma$) of the signal within the bandgap. A value $2\sigma$ above this $C_{g,av}$ was identified as the smallest value that is statistically different from the in-gap log(dI/dV). The energies at which the valence and conduction bands approach this value are used for linear fitting. Linear fits to the log(dI/dV) curve utilizing these points are shown in Fig. S5f. The crossing points between these linear fits and the $C_{g,av}$ are identified as the VBM and CBM respectively. The algorithm described above can identify band edge energies with a precision within the limit of quasiparticle lifetime broadening effects.[7]

**Discussion S3: Variation in the measured CDW gap**

The size of the CDW gap is extracted from STS scans as the energy difference between the two peaks on either side of $E_F$. A partial gap feature with two pronounced peaks around the Fermi level is consistently observed. In Fig. S5d,e, we show the observed tunneling spectra with different CDW gap values for different sample surfaces. Even for surfaces from the same crystal displaying identical CDW periodicities, we still observe large variations in the magnitude of the measured gap. In Fig. 1g, we present an average of STS spectra collected over a clean sample surface at 7 K, with an $2\Delta_{CDW} \sim 0.65$ eV, which falls near the middle of the 0.3 to 1 eV range. This anomalous variation in the STM measurements merits further attention. We find no correlation between the gap size and the presence of adatoms (Fig. S4f) (a signature of $Li^+$ intercalant concentration), which dispels the possibility of trivial ordering of intercalants leading to the observed order. Furthermore, we find no variation in the CDW periodicity in region with larger or smaller gap size, which shows that electronic heterogeneity does not affect the structural order.

We can compare these surface-sensitive, nm-scale STS measurements with our other probes of the CDW gap in **Li-CrSBr**, ARPES and infrared reflectivity (IR). ARPES measurements are similarly surface-sensitive but probe the sample on a μm-scale. The momentum-integrated ARPES data collected at 40 K show a clear peak at $E - E_F \sim -0.27$ eV. From this feature, we can estimate an average $2\Delta_{CDW} \sim 0.54$ eV, which falls near the middle of the range observed from STS measurements. Unlike STS and ARPES, IR probes the bulk (rather than the surface) of exfoliated flakes. At 100 K, the reflectivity shows a clear dip near 0.15 eV, which sets a lower bound on the magnitude of the gap. However, this signal should be dominated by excited carriers from sample regions with the smallest CDW gap; as such, this measurement cannot distinguish potential variation in the gap across the crystal. Indeed, while all three measurements are consistent with the opening of a partial gap by the CDW phase, these collective results appear to point to electronic heterogeneity that does not dramatically affect the periodicity of the CDW.

While we are unable to explicitly identify the source of this behavior, we can point to two factors that seem most likely to explain this effect. As noted in the text, our expectation is that intercalated lithium ions and THF molecules are disordered between CrSBr layers. While the bulk structure hosts a three-dimensional pocket to contain solvated cations, desolvated $Li^+$ species may sit closer to the electronically active Cr–S layers. Coulombic interactions with these "free" $Li^+$ ions could stabilize the CDW phase with no change in the overall doping level, which would be consistent with the observed data. Indeed, cation disorder has been used to explain heterogeneity in STM and STS data for other intercalated materials in the FeOCl structural family.[8] Beyond this cation disorder, strain may play a role in stabilizing (or destabilizing) the CDW in CrSBr. As shown in optical microscopy images of intercalated crystals and exfoliated flakes (and the large background enhancement in Raman spectroscopy), **Li-CrSBr** shows much rougher surfaces compared to the pristine material, likely due to the large interlayer expansion during intercalation. While we speculate that the electron filling dictates the periodicity of the CDW modulation, local strain can reasonably be expected to modulate the CDW gap. More intensive calculations would be needed to model the expected effects of disorder and strain on **Li-CrSBr**, and further scanning probe characterization may also help to explain the microscopic origin of this behavior.

The inhomogeneity in CDW gap size is also accompanied by large variations in the position of band edges. This could be an indication of locally varying tip-induced band bending which is a result of the insulating character of **Li-CrSBr** at 7K. Indeed, these variations are likely a result of local variations in the resistance, which could be correlated to inhomogeneous $Li^+$ doping or THF content.

**Discussion S4: Lithium-ion ordering imposing a structural order**

As noted in the main text, ordering of the intercalated lithium ions represents a trivial potential explanation for the superstructure observed in STM and SAED measurements. However, from the collective data, we believe this possibility can be effectively ruled out. First, we note that a superstructure of the lithium ions should necessitate an exact relationship between the lithium content and the periodicity of the superstructure ($q = xb^*$, where $x$ is the ratio of Li:Cr). Here, this relation would generate a periodic structure between 5 and 6 unit cells in length, rather than the experimentally observed periodicity of ~7 unit cells. While this mismatch seemingly precludes a superstructure tied to lithium ordering, we note that experimental error in the lithium content necessitates further evidence for cation disorder.

First, we point to the absence of a 3D supercell in bulk PXRD measurements. A large structural modulation induced by cation order should be reflected by the observation of a supercell in X-ray diffraction measurements. There are three possible explanations for the absence of these superstructure peaks: (1) The cation structure is intrinsically disordered between layers; (2) The small size of lithium ions prevents full determination of cation order; or (3) Lithium ions are ordered within the 2D layers but disordered along the *c* axis. While (2) would require neutron diffraction measurements more sensitive to small atoms, (3) can likely be ruled out by our collective data. If the unusually large magnitude of the CDW modulation (discussed in the main text) were caused by cation disorder, Coulomb effects should also favor intercalant order along the *c* axis. As such, in-plane intercalant order without out-of-plane order is unlikely in **Li-CrSBr**. Still, we cannot entirely rule out possibility (2) from XRD data alone.

Next, we point to observation of disordered surface species in STM topography measurements (Fig. S4e). The approximate density of these species ($\sim 3\times10^{12}$ cm$^{-2}$) is smaller than the expected density of intercalant species, consistent with the partial loss of surface species upon crystal cleavage and evacuation. The clear absence of any order for these species is consistent with the proposed intercalant disorder. Moreover, spatial variation in topographic heights observed across Fig. S4 would be consistent with variations in the LDOS tied to intercalant disorder and clustering. Indeed, similar behavior has been observed in STM studies on intercalated TiNBr and $MoS_2$.[9]

Collectively, these data point to disorder in the intercalant layers of **Li-CrSBr**, suggesting that trivial ordering of the intercalated species cannot explain the observed CDW features.

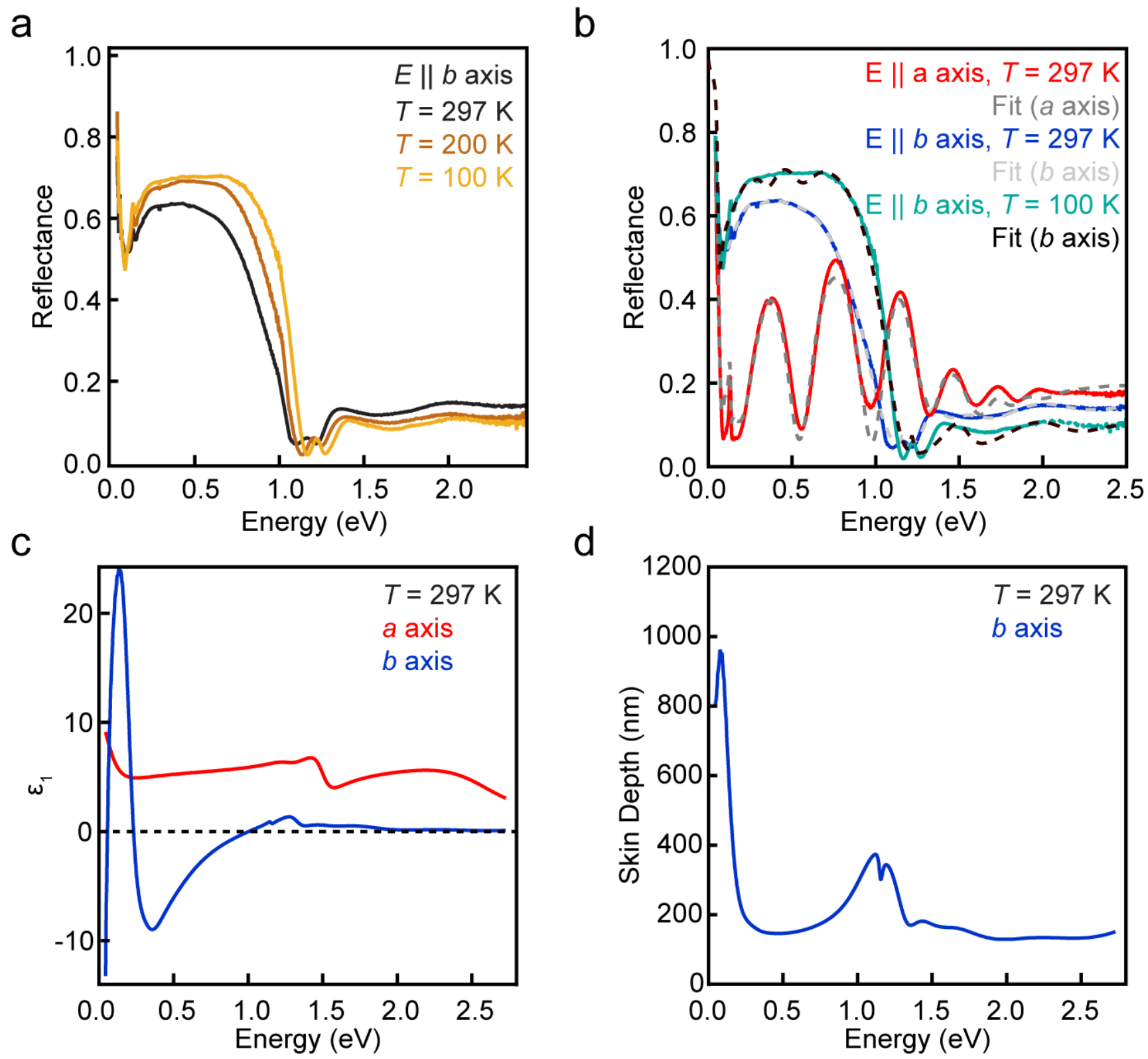

**Figure S6: Analysis of optical reflectivity data.**
a: Temperature dependent *b* axis optical reflectance of **Li-CrSBr**. As the temperature lowers from 297 to 100 K, the overall reflectance increases below the plasma edge (~1.2 eV) consistent with the metallic behavior along the *b* axis. b: Optical response along the *a* (297 K) and *b* (100 and 297 K) axes with fittings shown in dashed lines. At $T$ = 297 K for $E \parallel b$, we can extract the plasma frequency of 4.055 eV, which is then used to calculate the carrier density noted in the main text.[5,10] We note that **Li-CrSBr** exhibits electronic anisotropy similar to other quasi-1D CDW materials, including molybdenum blue bronze ($K_{0.3}MoO_3$), $NbSe_3$, and $(TaSe_4)_2I$.[11,12,13] c: The extracted real part of the complex dielectric function ($\varepsilon = \varepsilon_1 + i\varepsilon_2$) for *a* and *b* axes of **Li-CrSBr**. Compared with the dielectric response in the *a* axis (red), the *b* axis response is metallic with a zero-crossing of ~ 1 eV and at the low energy limit. d: The extracted skin depth for light polarized along the *b* axis at 297 K. Because the skin depth of the optical measurements is above 100 nm across the entire energy range, this low-energy feature in the reflectivity is consistent with the presence of a bulk CDW with a partial gap near $E_F$.

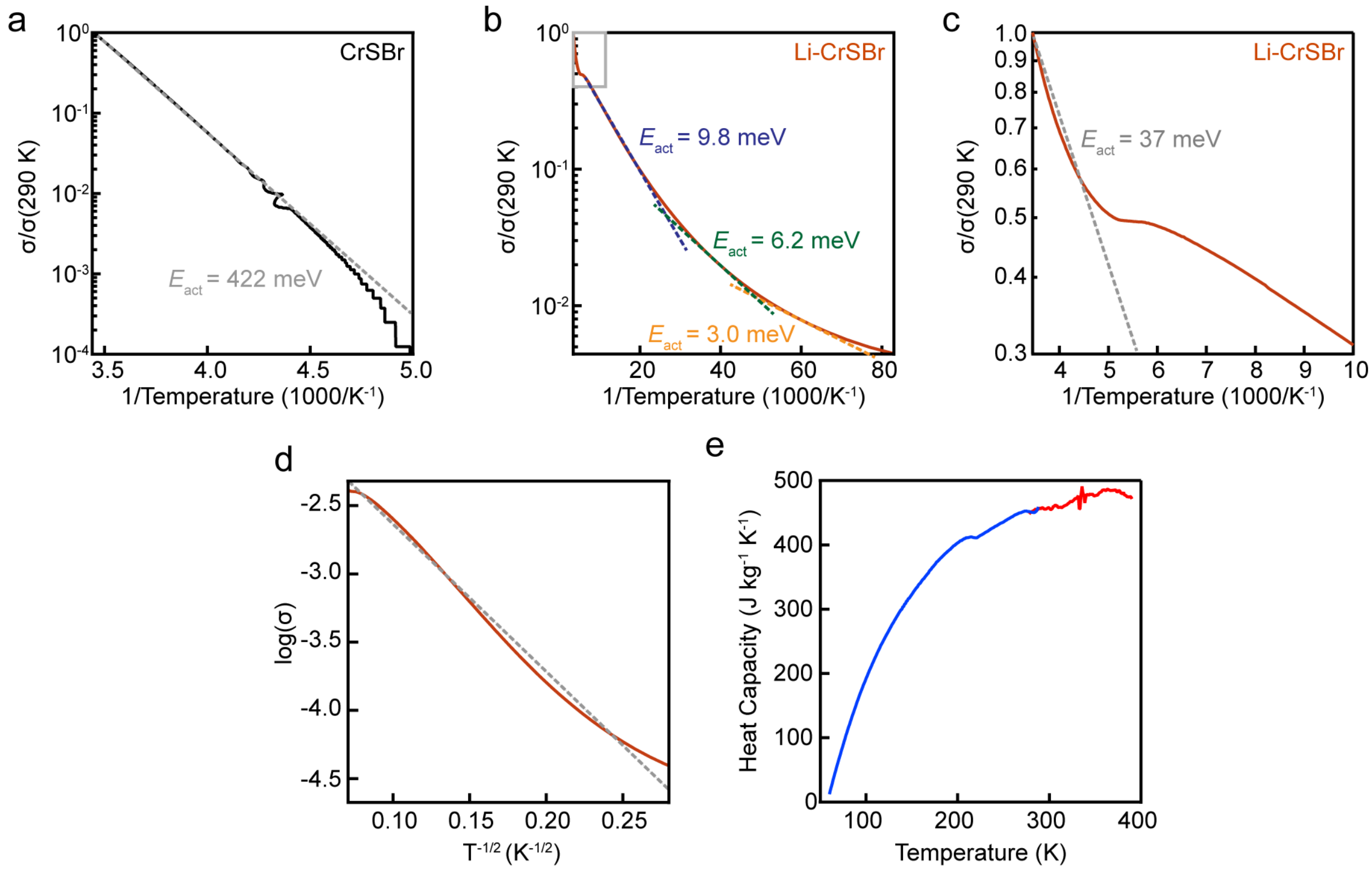

**Figure S7: Analysis of electronic transport and heat capacity.**
a,b,c: Arrhenius fitting for CrSBr (a) and **Li-CrSBr** (b,c) using normalized conductance. The high-temperature behavior of **Li-CrSBr** is boxed in (b) and magnified in (c) for clarity. Activation energies for CrSBr at all measurable temperatures and **Li-CrSBr** are given. While high-temperature data (200−300 K) is nearly linearly for **Li-CrSBr**, low temperature data shows decreasing activation energies with decreasing temperatures, as listed. d: Low-temperature behavior is fit to a Efros–Shklovskii variable range hopping model with a deviation from linearity for **Li-CrSBr**. e: Heat capacity of **Li-CrSBr**, where one sample was measured below 290 K (blue) and another sample was measured above 280 K (red). Due to uncertainly in mass, the high-temperature data was adjusted to overlap with the low-temperature data within the mass error. The data shows no sharp features in the range 60−390 K. There is a broad anomaly at 215 K in heat capacity, overlapping with the onset of magnetic order rather than charge order. This type of feature is also present in pristine CrSBr overlapping with its $T_N$.[14] The small feature at ~330 K is an artifact of the sample and is too small to be a CDW transition.

**Discussion S5: The theoretical model**

Assuming weak interactions with intercalated cations and weak interlayer interactions (as suggested by the large interlayer spacing), we can effectively treat **Li-CrSBr** as individual layers of CrSBr with electrons doped into the conduction bands. To model this system, we localized the four lowest energy unfilled bands of a ferromagnetic monolayer of CrSBr (modeling a monolayer effectively mimics the large interlayer spacing of **Li-CrSBr**). This minimal model includes two Cr atoms per cell, each with one Wannier orbital (Fig. 3a). Among these orbitals, there are four key parameters: one intra-chain hopping parameter (t), one intra-chain Coulomb interaction ($V^{\mathrm{intra}}$), and two inter-chain interactions ($V_1^{\mathrm{inter}}$ and $V_2^{\mathrm{inter}}$).

These parameter values are derived from first-principles density functional theory (DFT) and constrained random phase approximation (cRPA) projected onto the Wannier orbitals. The entire system comprises periodically repeated quasi-1D chains, coupled via Coulomb interactions. The model's validity is evident from the hopping matrix elements among the Wannier orbitals (see Table S2), where the intra-chain hopping ($t$ = −0.481 eV) is the only significant element, highlighting the material's quasi-1D nature. To keep the model simple, we retained three Coulomb interaction parameters and ignored the long-range ones. Including long-range interactions can lead to a more complex phase diagram, which we report separately.[15]

Using this model, we calculated the electronic structure via the periodic Hartree-Fock method. The periodicity depends on the number of doped electrons, affecting the quasi-1D Fermi momentum $k_F = \pi n_{\mathrm{elec}}/b$, where $b$ is the lattice size along the chain direction. Given that the number of doped electrons in experiments is ~0.14−0.18/Cr, the simulation agrees well with the STM pattern. We further examined the influence of the second inter-chain Coulomb interaction ($V_2^{\mathrm{inter}}$), as shown in Fig. S8c. $V_2^{\mathrm{inter}}$ is unfavorable for forming a CDW state with a phase π, as the Coulomb interaction between the second neighboring chains starts to dominate after $V_2^{\mathrm{inter}} > 1.2$ eV. The value of $V_2^{\mathrm{inter}}$ from the *ab initio* calculations falls within the region where the CDW state is more stable.

**Table S4: Model hopping parameters (*t*) and Coulomb interactions (*V*).**
The values in red are used in the minimal model. *d* indicates distance.

| $d_{\{ij\}}$ [Å] | $t_{\{ij\}}$ [eV] | $V_{\{iijj\}}$ [eV] | comment |
|---|---|---|---|
| 0.000 | −0.831 | 1.875 | on-site |
| 3.513 | 0.034 | **0.742** | 2nd neighbor inter-chain |
| 3.571 | 0.093 | **0.804** | 1st neighbor inter-chain |
| 4.746 | **−0.481** | **0.834** | intra-chain |

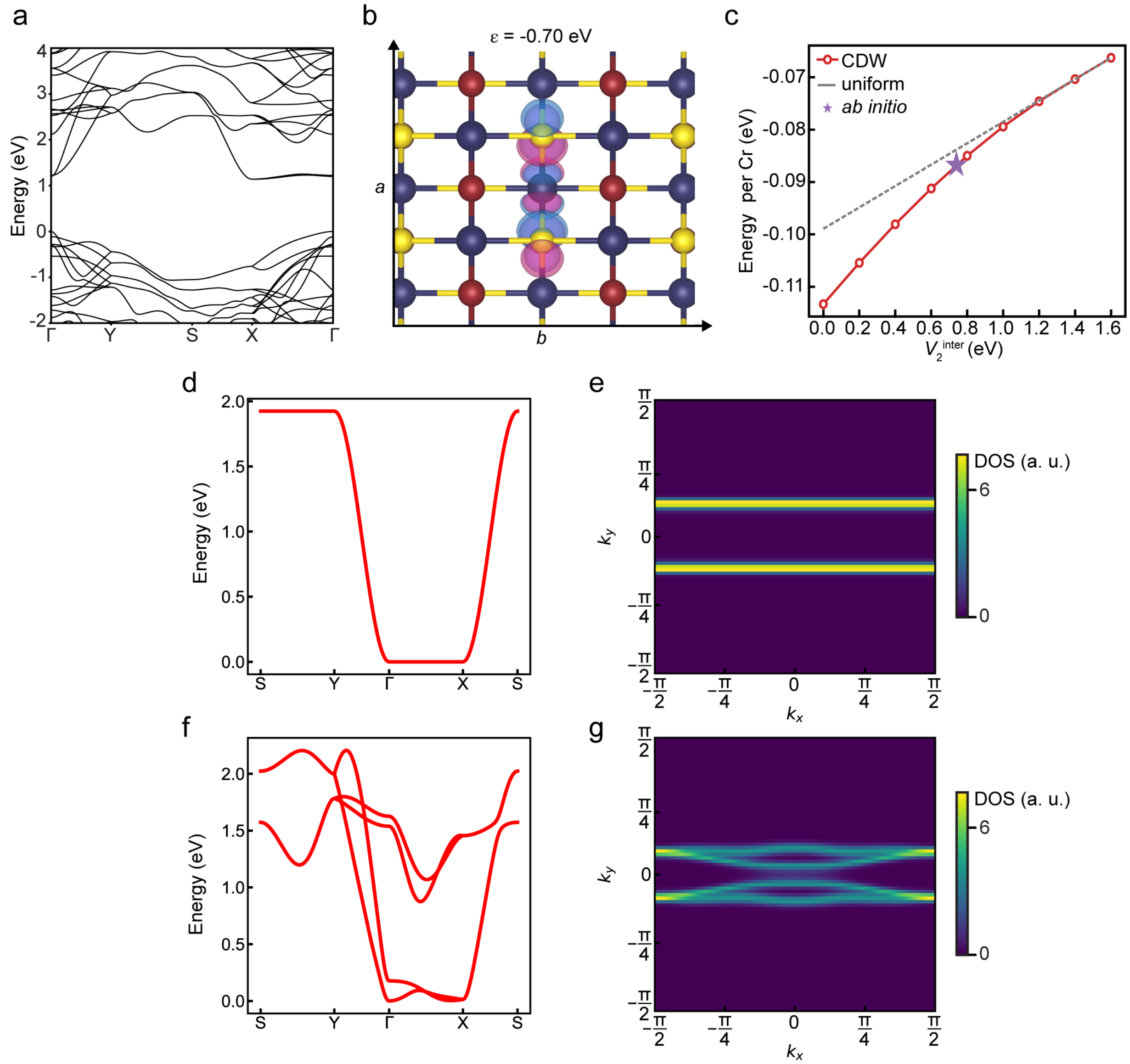

**Figure S8: Calculated band structures and details of the minimal model.**

a: Full band structure for CrSBr from DFT+U ($U$ = 4 eV, $J$ = 1 eV). b: Maximally localized Wannier functions for the $d_{x^2-y^2}$-like conduction bands of CrSBr. c: The influence of the second nearest-neighbor interchain Coulomb interaction ($V_2^{\text{inter}}$) on energy. d, e: Band structure (d) and Fermi surface (e) of the 2-band minimal model used for the minimal model. There are no hopping matrix elements between the two chains of Cr, so the bands perfectly overlap. f, g: Band structure (f) and Fermi surface (g) of the 4-band full model to show imperfect nesting of the Fermi surface in **Li-CrSBr**.

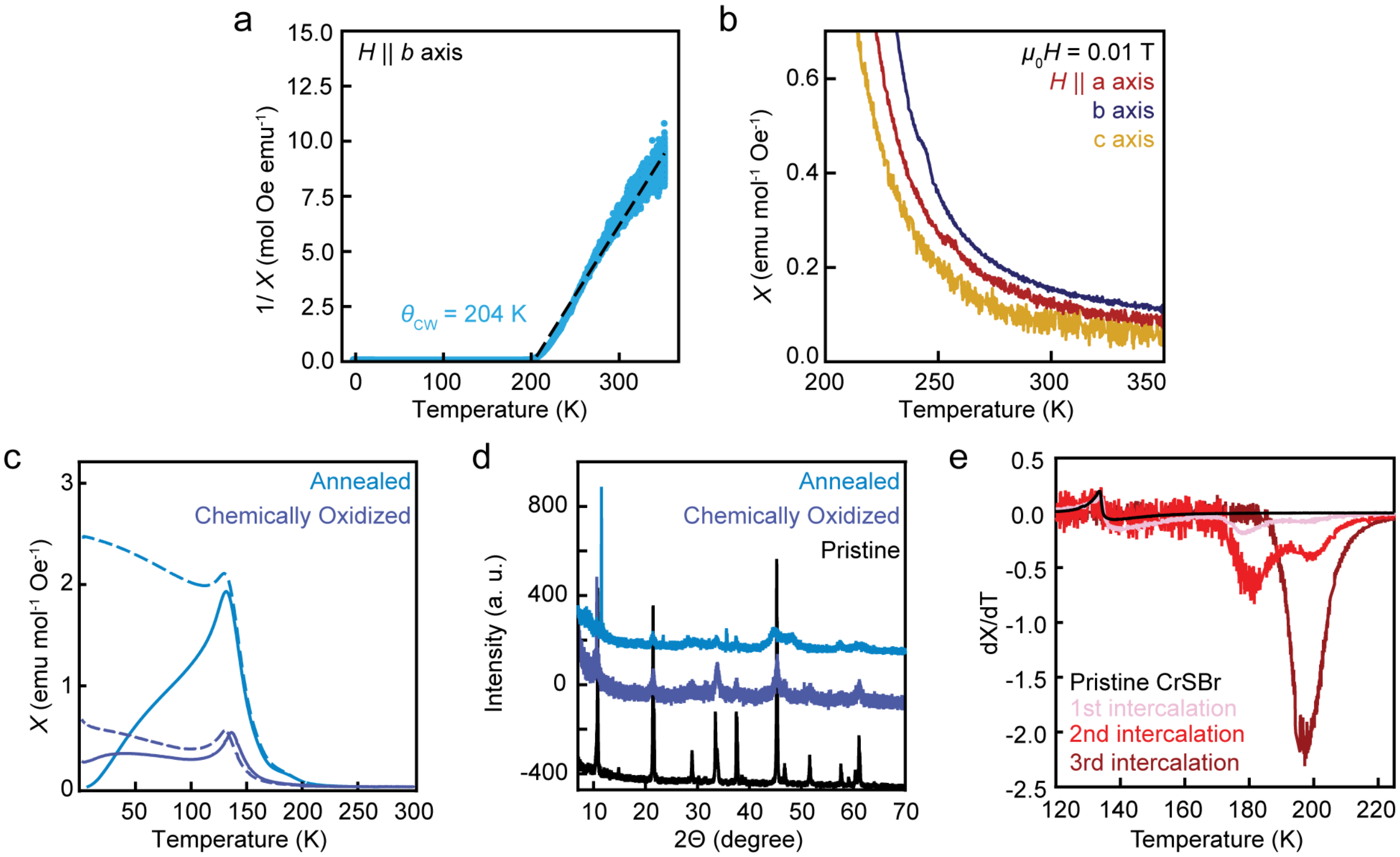

**Figure S9: Analysis of magnetic data, de-intercalation, and partial intercalation.**
a: Curie-Weiss plot of **Li-CrSBr** when $H$ || $b$ axis ($H$=100 Oe) to show the Curie-Weiss temperature ($\theta_{CW}$) after a diamagnetic correction was taken. Curie-Weiss fitting is done at the high-temperature region in the range 230−350 K. b: Magnification of the high-temperature temperature-dependent susceptibility ($\chi$) of field-cooled **Li-CrSBr** along all three crystallographic axes. c: $\chi$ of **Li-CrSBr** after annealed at 200°C in an inert atmosphere (light blue) and after chemical oxidation with ferrocenium hexafluorophosphate in THF in an inert atmosphere (dark blue) where $H$ || $b$ axis and $H$=100 Oe. ZFC is shown in with a solid line; FC is dashed. Both show negligible features at 200 K and a prominent feature at 132 K, indicating the reemergence of pristine CrSBr and reversibility of intercalation. d: PXRD of annealed, chemically oxidized and pristine CrSBr to show the general alignment of the Bragg peaks, especially of the 001 peak at 10.7°. e: Derivative of $\chi$ with respect to temperature versus temperature as intercalation progresses. First intercalation (pink) shows a small FM feature at ~170 K; second intercalation (red) shows two FM features at ~170 K and 200 K; third intercalation (dark red) shows 1 strong feature at 200 K indicating full intercalation. These magnetic transitions appear at specific temperatures, suggesting that the magnetic ordering temperature does not shift continuously with increasing Li content, and instead progresses through discrete stages of intercalation. We speculate that these stages may correspond to CDW phases with larger periodicities than what we observe here, though we cannot rule out more conventional intercalation staging (with different $c$ axis periodicities) as an explanation for this effect.

**Discussion S6: Saturation Magnetization**

As noted in the main text, the saturation magnetization for **Li-CrSBr** of 4 $\mu_B$ exceeds the spin-only value we would expect for this material (3.17 $\mu_B$), assuming fully spin-polarized conduction bands. For the local $C_{2v}$ symmetry of each Cr ion, the degeneracy of the 3*d* orbitals is fully broken; consequently, any orbital contribution from the localized Cr moment should be small and is unlikely to explain the different between the expected and experimental value. Similarly, mass error (even on the order of 5−10%) cannot explain the observed difference.

Focusing instead on the paramagnetic phase, Curie-Weiss analysis yields a Curie constant of 15.7 emu K $mol^{-1}$ $Oe^{-1}$, dramatically larger than the expected value of 2.06 emu K $mol^{-1}$ $Oe^{-1}$ for non-interacting spins in this mixed-valent $d^3/d^4$ material. While the enhanced value at high temperatures could be tied to ferromagnetic clusters of local moments, as observed in some mixed-valent transition metal oxides, it may also signify an effect of the CDW phase on the magnetism of this material, both above and below $T_C$.[16] Indeed, anomalous effects of a CDW phase on the magnetization have been observed in FeGe, which also hosts co-existing CDW and magnetic orders, though further experiments are needed to support and understand the possible link between these phases in **Li-CrSBr**.[17]

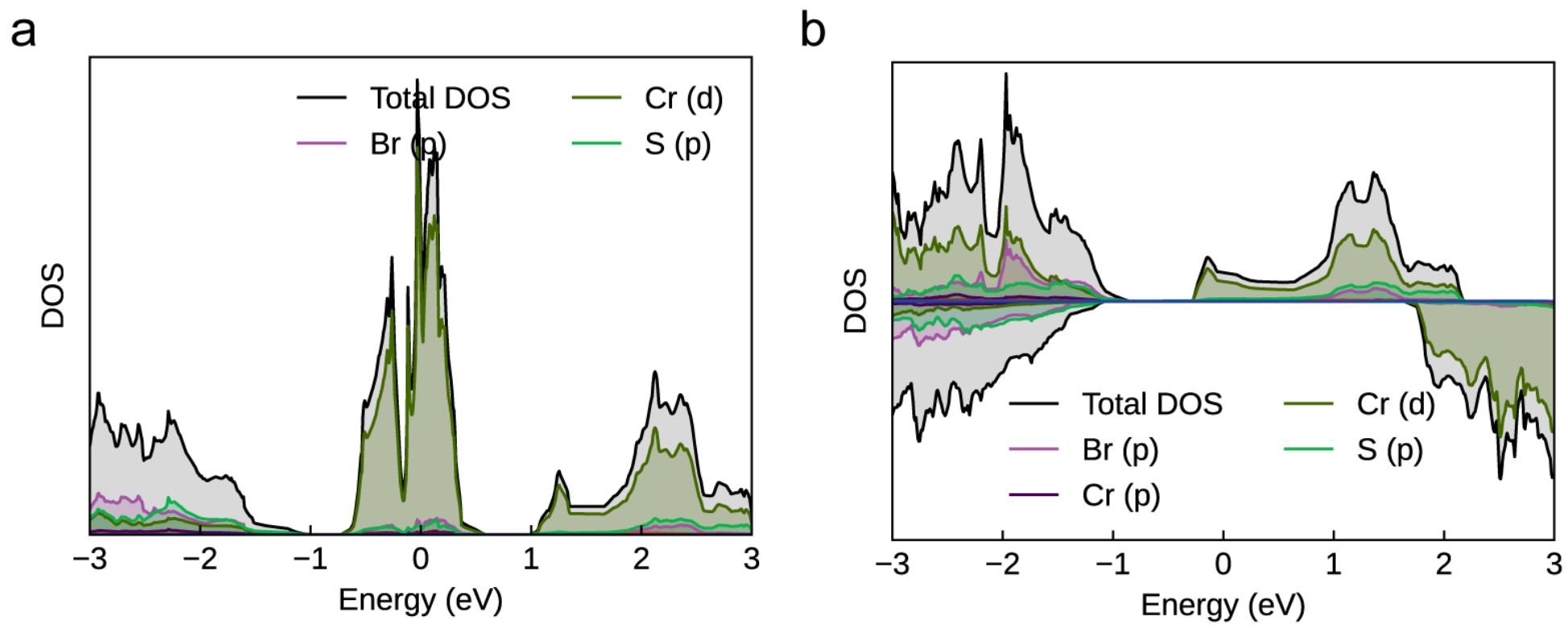

**Figure S10: Density of states calculations for doped CrSBr.**
a,b: Calculated first-principles DOS for bulk CrSBr doped with 1/7 electrons per Cr atom in the paramagnetic (a) and ferromagnetic (b) states. In the FM state, exchange splitting does not push any DOS between −1 to 0 eV. We note this is calculated in the charge uniform, not CDW, state, though this should not affect any analysis for features away from $E_F$.